\documentclass[acmsmall]{acmart}

\AtBeginDocument{%
  \providecommand\BibTeX{{%
    \normalfont B\kern-0.5em{\scshape i\kern-0.25em b}\kern-0.8em\TeX}}}

\setcopyright{acmcopyright}
\copyrightyear{2018}
\acmYear{2018}
\acmDOI{10.1145/1122445.1122456}

\acmJournal{JACM}
\acmVolume{37}
\acmNumber{4}
\acmArticle{111}
\acmMonth{8}
\usepackage{multirow}
\usepackage{caption}
\usepackage{subcaption}
\usepackage{amsmath}
\usepackage{footnote}

\usepackage[T1]{fontenc}
\PassOptionsToPackage{ruled}{algorithm2e}
\usepackage{color}
\usepackage{array}

\usepackage{textcomp}
\usepackage{multirow}
\usepackage{algorithm2e}
\usepackage{amsbsy}
\usepackage{graphicx}

\begin{document}
\title{Learning Explicit and Implicit Latent Common Spaces for Audio-Visual Cross-Modal Retrieval}


\author{Donghuo Zeng}
\affiliation{\institution{KDDI Research, Inc.}\streetaddress{2-1-15 Ohara}\city{Fujimino}\state{Saitama}\postcode{356-8502}\country{Japan}}

\author{Jianming Wu}
\affiliation{\institution{KDDI Research, Inc.}\streetaddress{2-1-15 Ohara}\city{Fujimino}\state{Saitama}\postcode{356-8502}\country{Japan}}

\author{Gen Hattori}
\affiliation{\institution{KDDI Research, Inc.}\streetaddress{2-1-15 Ohara}\city{Fujimino}\state{Saitama}\postcode{356-8502}\country{Japan}}

\author{Yi Yu}
\affiliation{\institution{National Institute of Informatics, SOKENDAI}\streetaddress{2-1-2 Hitotsubashi}\city{Chiyoda-ku}\state{Tokyo}\postcode{100-0003}\country{Japan}}
\author{Rong Xu}
\authornote{He is a collaborator of Yi Yu.}
\renewcommand{\shortauthors}{Trovato and Tobin, et al.}

\begin{abstract}
Learning common subspace is prevalent way to solve the problem of data from different modalities having inconsistent distributions and representations that cannot be directly compared when achieved in cross-modal retrieval.
Previous cross-modal retrieval methods focus on projecting the data between modalities into a common latent subspace by learning the correlation between them to bridge the modality gap. However, due to the rich semantic information in the video, the heterogeneous nature of audio-visual data leads to heterogeneous gaps in the distribution of modalities, which are difficult to be bridged into a common subspace by the previous cross-modality learning methods, which may lead to the loss of important semantic content in the video when the modality gap is eliminated by supervised learning, e.g., optimizing pairwise correlation, while the semantics of the categories may undermine the properties of the original features In this work, we aim to learn effective audio-visual representations to support audio-visual cross-modal retrieval (AVCMR). We propose a novel model that maps audio-visual modalities into two distinct shared latent subspaces: explicit and implicit shared spaces. In particular, the explicit shared space is used to optimize pairwise correlations, where learned representations across modalities capture the commonalities of audio-visual pairs and reduce the modality gap. The implicit shared space is used to preserve the distinctive features between modalities by maintaining the discrimination of audio/video patterns from different semantic categories. Finally, the fusion of the features learned from the two latent subspaces is used for the similarity computation of the AVCMR task. The comprehensive experimental results on two audio-visual datasets demonstrate that our proposed model for using two different latent subspaces for audio-visual cross-modal learning is effective and significantly outperforms the state-of-the-art cross-modal models that learn features from a single subspace.

\end{abstract}

\begin{CCSXML}
<ccs2012>
 <concept>
  <concept_id>10010520.10010553.10010562</concept_id>
  <concept_desc>Computer systems organization~Embedded systems</concept_desc>
  <concept_significance>500</concept_significance>
 </concept>
 <concept>
  <concept_id>10010520.10010575.10010755</concept_id>
  <concept_desc>Computer systems organization~Redundancy</concept_desc>
  <concept_significance>300</concept_significance>
 </concept>
 <concept>
  <concept_id>10010520.10010553.10010554</concept_id>
  <concept_desc>Computer systems organization~Robotics</concept_desc>
  <concept_significance>100</concept_significance>
 </concept>
 <concept>
  <concept_id>10003033.10003083.10003095</concept_id>
  <concept_desc>Networks~Network reliability</concept_desc>
  <concept_significance>100</concept_significance>
 </concept>
</ccs2012>
\end{CCSXML}

\ccsdesc[500]{Computer systems organization~Embedded systems}
\ccsdesc[300]{Computer systems organization~Redundancy}
\ccsdesc{Computer systems organization~Robotics}
\ccsdesc[100]{Networks~Network reliability}

\keywords{Explicit common subspace, implicit common subspace, CCA layer, audio-visual cross-modal learning}
\maketitle

\section{Introduction}
\begin{figure}%
    \centering 
    \includegraphics[width=0.8\linewidth]{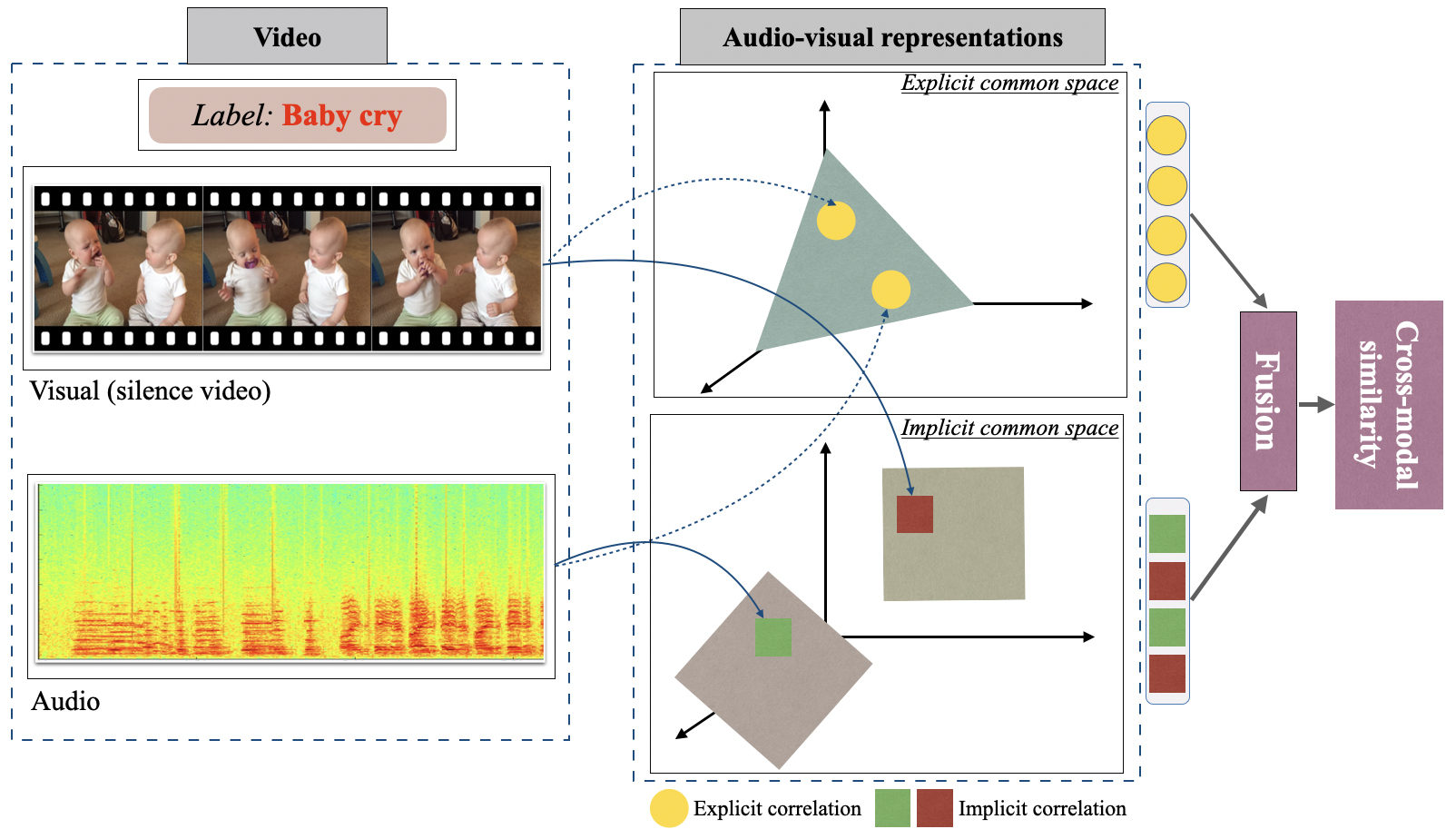}%
    \caption{Audio and visual are two main tracks from annotated video, the relation between audio and visual is captioned by representations learning method via projecting the audio-visual features into two distinct subspaces: explicit and implicit common subspaces to learn new representations, respectively. At last, the fusion features of these learned representations are utilized for audio-visual cross-modal similarity calculation.}%
    \label{fig:subspaces}%
\end{figure}

With the development of the Internet and the gradual improvement of the functions of mobile devices, users are progressively becoming self-publishing consumers in recent years, short video applications such as TikTok, Instagram, Snapchat have been users' favourite entertainment tools, some people almost cannot live without short videos. In fact, videos have become the next generation of artificial intelligence, the rich semantics provide us with challenges and opportunities to extract useful information and build interaction between different modalities. Cross-modal learning methods have found success at the intersection of computer vision and natural language processing. However, as recently published papers~\cite{zhen2019deep, roy2020zscrgan} show, there is still room for improvement in some common image-text benchmark datasets, such as the Wikipedia dataset~\cite{pereira2013role} and the NUS-WIDE-10k dataset~\cite{chua2009nus}.

Moreover, there are few research works and available datasets on audio-visual cross-modal learning, which should be taken into account. In recent years, some works have applied audio-visual joint embedding learning to some tasks, such as Multimodal Sentiment Analysis~\cite{hazarika2020misa}, audio-visual separation~\cite{lu2018listen}, Visual2sound generation~\cite{zhou2018visual}, and audio-visual cross-modal retrieval~\cite{zeng2020deep}. The AVCMR task faces the same challenge as other cross-modal learning methods, since the data from different modalities cannot be directly compared. In addition, audio-visual cross-modal learning is more challenge and potential than other combination of cross-modal learning such as image-text because the research situation of audio-visual cross-modal learning is still in its infancy.

A general approach to bridge the heterogeneous gap of two different modalities is by representation learning method~\cite{Zhu_2019_CVPR, ter8771379, wang2017adversarial, zeng2019learning}, where a common subspace is learned to obtain the same dimension of the representation in all modalities by optimizing the correlation between them, so that the similarity of the new representations can be measured directly. In the past, several cross-modal learning methods have been reported that use different methods to project the cross-modal data into a common subspace. \textcolor{red}{Traditional methods such as CCA-variant~\cite{hardoon2004canonical, rasiwasia2014cluster, zeng2020deep} learning linear/non-linear projections by optimizing pairwise correlation to generate a pair of new cross-modal features, some CCA-variant~\cite{hardoon2004canonical,zeng2020deep} have success in incorporating deep Learning methods to learn high-level semantic representation. For this reason, we adopt the idea of combining pairwise correlation learning and deep learning methods in the explicit common subspace. }

Audio and visual are different modalities that are extracted from the same video, moreover, they share the same motive and goal of semantics that allows us to learn the correlation between them by cross-modal learning methods. To ensure the correlation do not deviate from the original semantic information such as human annotation. Unlike the YouTube-8M~\cite{abu2016youtube}, VGG-SOUND~\cite{chen2020vggsound}, and Kinetics~\cite{will33824} dataset are labeled by computer vision techniques, in our work, we use the audio-visual datasets including VEGAS~\cite{zhou2018visual} and AVE~\cite{tian2018audio} with audio-visual human double-check labels. \textcolor{red}{Compared with the previous works~\cite{wang2016effective, wang2017adversarial}, the category information is only used to learn individual features for each modality or to learn from the inter-modalities, which may cause the semantic information to be lost during the joint subspace learning. In this case, we learn the semantic information in an additional shared subspace without explicit learning of a shared representation in the implicit shared subspace. To avoid that the two learned joint subspaces learn from different directions, which leads to the inability to merge the learned representation. The implicit shared subspace is the supplement of explicit shared subspace, which is unsupervised learning is based on the concept of mutual
information~\cite{borga1998learning}. Our proposed model applies a constraint loss to control the consistency of the two shared subspaces. Finally, the end representation is fused by the features mapped from two subspaces. }

The main contributions of can be summarized as follows: We propose a simple and flexible audiovisual cross-learning model that extracts a comprehensive and unmixed representation across modalities by learning from the explicit shared space and the implicit shared space. After merging the output representations from two subspaces, the computation of cross-modality similarity is promoted. Experiments on the VEGAS and AVE datasets show that learned representations with simple fusion can outperform state-of-the-art models.

The rest of this paper is organized as follows. Section~\ref{relatedwork} describes related work in common spaces of cross-modal learning. Section~\ref{approachsection} explains our approach and evaluates it experimental explanation in Section~\ref{experimentsection}. Section~\ref{results&analysis} presents the results and conducts an analysis. Section~\ref{conclusion} concludes this paper.

\section{Related Work}
\label{relatedwork}
This section predominantly discusses the related works of common subspace learning in two aspects and analysis the difference of our proposed model compared with these related works.

Since the heterogeneous gap between modalities limits the similarly computation of parings, existing cross-modal learning methods~\cite{wang2016comprehensive, muller2018cross, kaur2021comparative} aim to learn a common space to generate modality gap-free representations so that eventually the transformed data can be compared using distance metrics, such as cosine similarity distance and Euclidean distance. Many models~\cite{ayyavaraiah2019cross} have been proposed to learn such a shared space, and it is difficult to summarize these approaches with well-defined categories. In previous work on the survey of cross-modal retrieval~\cite{peng2017overview, priyanka2013analysis, ayyavaraiah2018joint, peng2017cross}, the definition of categories depends on whether the learning space is a real-valued space or a binary/Hamming space. In this paper, we divided the related works in two lines by the cross-modal similarity calculation of the model is with explicit common representation or not. 

\subsection{Explicit Common Subspace Representations Learning}
Explicit common subspace representations are obtained by learning pairwise correlations to calculate the cross-modal similarity.

\subsubsection{Canonical Correlation Analysis (CCA) variant methods} 
CCA~\cite{hardoon2004canonical} and CCA-variant~\cite{akaho2006kernel, andrew2013deep, rasiwasia2014cluster, ranjan2015multi, yu2018category, zeng2020deep, dorfer2018end} methods are typical methods to learn explicit common subspace for generating transferred features for each modality. The CCA~\cite{hardoon2004canonical} is to find linear transformation for each modality to optimize the correlation of the transferred data in the common space. The KCCA~\cite{akaho2006kernel} extends CCA with a kernel trick by projecting the data into high dimension space to generate non-linear representations. The DCCA~\cite{andrew2013deep} extend CCA with deep neural network skills to learn complex nonlinear transformations of cross-modal data so that the generated representations are highly linearly correlated. The Cluster-CCA~\cite{rasiwasia2014cluster} is a cluster-based CCA method that is to caption the possible corresponds within cluster to learn the linearly correlation of the transferred data. Compared with existing CCA-variant methods, The Multi-label CCA~\cite{ranjan2015multi} is other way to extend CCA method, which learns common space to extract high-level semantic representations by the multi-label information instead of learning explicit pairwise correlation. The C-DCCA~\cite{yu2018category} is a significant extension of DCCA model, which learns correlation of data from two modalities for specific single label annotation. The TNN-C-CCA~\cite{zeng2020deep} applies audio-visual learning based task specific loss function to improve the Cluster-CCA method, which is not only take the pairings correlation into account, but also consider the non-pairings correlation, by exploiting triplet loss to update the weights of triplet neural networks for the constructed triplet inputs. The NLCCA method~\cite{hsieh2000nonlinear} is to extend CCA methods via using three feedforward neural networks, the first neural network consists of two branches that is to maximize the correlation between the two output representations. The other two neural networks use to project the canonical variables into the two original set of data. To enhance the correlation of pairings in the of CCA mechanism, the $CCAL-L_{rank}$~\cite{dorfer2018end} employ the pairwise ranking losses to reform the performance of CCA methods, by learning better embedding with the combination of CCA layer with present objectives for embedding space learning. In the application of the~$CCAL-L_{rank}$, task specific loss functions can be set on the output of final layer. The ICCA~\cite{ShaoZSY17} method is toward improving CCA for cross-modal learning tasks, by solve the essential problems of CCA methods including the weakness of intra-modal semantic consistency ignored, the troublesome non-linear correlation learning, and similarity measure indirectly optimized. In addition, CCA can be applied in improving the adversarial learning method such as GAN, this work~\cite{harada2019biosignal} argues that the input and generated data is not clear and the generated data is out of control in GAN, by using CCA to control the generated samples with optimizing the canonical correlation between input and the generated data during the training at the same time.

\subsubsection{Deep Neural Networks (DNN) Based CMR methods} With the success of DNN models, which allows to extend to the multimodal learning by supplying scalable nonlinear transformations for valid cross-modal representation learning~\cite{ngiam2011multimodal, CaoLWYY16, yang2017pairwise, carvalho2018cross, zhang2016pl, KarpathyJL14,LiLLZYZ20, MenonSC15, JiangWLZLTZ15,  socher2014grounded,  FromeCSBDRM13,  Geigle2021RetrieveFR}.
%unfinished......
Deep learning has already succeed in the representation learning for single modality, such as text, image, and audio, bridging the heterogeneous gap across modalities is still a big challenge for deep learning methods.
Multimodal Deep Learning~\cite{ngiam2011multimodal} introduced a representation learning for multimodal tasks, in particular, this cross-modal representation learning approach demonstrate if multiple modalities data used to learn can get more powerful features. The DVSH~\cite{CaoLWYY16} is a hybrid visual-semantic fusion network that use to capture the intrinsic image-text correspondences for learning joint embedding space of these two modalities, in particular, the inputs of fusion network are two modality-specific hashing networks to generate perfect hashing codes. This work~\cite{carvalho2018cross} propose a unified framework that takes both the pairwise similarity for leveraging retrieval and class-based features in a common space, they applied double triplet loss to take the the high-level semantic and the fine-grained features into account. DeViSE~\cite{FromeCSBDRM13} is to solve the problem of large numbers of object categories that is hard to recognise without enough training data, by introducing a combination model encompassing visual and semantic embedding learning with image data and its unstructured text information. The cooperative and joint methods for promoting the CMR~\cite{Geigle2021RetrieveFR} proposed a novel framework for the limitation of current CMR methods less scalable and huge retrieval cost, by combining a twin networks to encode all data for efficient initial retrieval and a cross-encoder structure for retrieved same unit of data. The $C^{2}MLR$~\cite{JiangWLZLTZ15} model is to discover better representations and optimize pairwise rank function by enhancing both local alignment and global alignment, in detail, local alignment is about the alignment within the visual object and textual words, and global alignment is about the visual high level and semantic high level alignment. Similar to this work, the Deep Fragment~\cite{KarpathyJL14} present a model, unlike other methods that projected cross-modal representations into a common space, consider the finer level information, visual objects, and sentence fragment. Except the rank function during mapping the data into common space, the work also added a new fragment alignment function to learn the alignment between these fragments. The SDPP~\cite{LiLLZYZ20} extract the correlation in a common space by additionally eliminating the modality-specific information which is not related to the cross-modal retrieval task, with Hilbert Schmidt Independence Criterion (HSIC) method. A Pairwise Classification Approach~\cite{MenonSC15} for CMR is to retrieve the best matching in the retrieval candidate list when given a query from other modality based on the pairwise classification, which is not only can learn the correlation on positive links between two entities, but also can caption the unlabeled links in the cross-modal retrieval datasets. The DT-RNNs~\cite{socher2014grounded} designs dependency trees structure to project the sentences into a space to retrieve related images with these sentences, DT-RNNs learn the action and agents information in the sentence that can learn better representation from the logic of word sequence and the language expression. The PRDH~\cite{yang2017pairwise} work argued that existing cross-modal hashing method directly using manual features and overlook the internal correlation that hurt the performance of the CMR task. The PRDH is an end-to-end deep learning method that can capture the internal correlation between cross-modal data. The PL-ranking~\cite{zhang2016pl} is a low-rank optimization model that consider the top samples and improve the precision of the top samples in the rank list for a given query when learning representation in the low-dimension shared spaces. PL-ranking used pairwise ranking loss for the top of rankings and considered the classification information by a applying a list-wise constraint, in the end, the correlation between data features and label embedding is enhanced by a low-rank regularization.

\subsubsection{Generative Adversarial Networks (GAN) Based CMR Methods}
Currently, generative adversarial networks (GAN)~\cite{roy2020zscrgan, chaudhuri2020simplified, xu9296975, Zhu_2019_CVPR, ShangZZS19} prove a large potential room can improve the performance of cross-modal learning. The multi-stream encoder-decoder model~\cite{chaudhuri2020simplified} not only ensures enhancing the correlation between the RGB and sketch image, but also guarantee the semantic feature is discriminative in encoded feature space. In other word, the model can preserve the semantic space while eliminating the modality dependency during the training process. 
The ZSCRGAN~\cite{roy2020zscrgan} is to address the problem of previously unseen classes in query on training set by presenting a GAN-based zero-shot learning model for text to image retrieval. When given a text as the query, the model can get relevant images with zero-shot learning by using an Expectation-Maximization. The DLA-CMR model~\cite{ShangZZS19} applied dictionary learning algorithm to reconstruct discriminative representations, and adversarial learning exploit statistical characteristics of each modality. The model maintains the specificity of training/testing set, and ensures the weight of main features increases while other features decreases. The JFSE model~\cite{xu9296975} investigated the problem of adversarial learning in existing GAN-based cross-modal methods into two kinds and introduced Joint Feature Synthesis and Embedding method that combined the multimodal feature and representation learning in common space to remedy the investigation problem by developing a conditional wassertein GAN modules for the cross-modal input data.
The R2GAN model~\cite{Zhu_2019_CVPR} is short for a new version of GAN, named Recipe Retrieval Generative Adversarial Network, which is to study the feasibility of the retrieval problem that generates image from text. The R2GAN model consists of two discriminative models and one generative model where achieved with a two kinds of rank loss in both spaces.

\subsubsection{Semantic Supervised Learning Based CMR Methods}
With the growth of data scale, the accuracy of label annotation gets hard, to overcome the situation, it requires us to extract more latent useful features. One common solution~\cite{ter8771379, xie2020multi, XuLZ20} for this case is to utilize the label feature as an additional feature to aid cross-modal learning.

The TANSS model~\cite{ter8771379} is motivated by the zero-shot cross-modal retrieval that there are some categories in the test set that have not appeared in the training set. The TANSS model is via designing three subnetworks to address the zero-shot problem in CMR tasks. In detail, the first two subnetworks are to learn the internal information inside the data for each set while keeping the instances across modalities sharing the same semantic information. The third subnetwork is a self-supervised learning method that is used to transfer the existing labels to unseen labels in the test. In the SCCMR~\cite{XuLZ20}, they explained the shortcoming of the existing supervised cross-modal retrieval including overlooking the abundant semantic information without annotations during bridge the heterogeneous gap. The SCCMR model is to globally optimize the combination of label prediction and the projection matrices, while exploiting the manifold graph structure to mine the latent semantic information across modalities. 

In the cross-modal hashing methods, semantic inconsistency problem after bridging the heterogeneous gap is also a big challenge for CMR task. The (CPAH) model~\cite{xie2020multi} tried to completely preserve the semantic consistency during the correlation learning between two sets by introducing a consistency refined mechanism to depart the different modalities data into two different latent spaces, then using an adversarial learning to ensure the representation from shared space full of semantic consistency and generate compact and discriminative representations in the end.

\subsection{Implicit Common Subspace Representations Learning}
Accept existing heterogeneous gap problem in cross-modal learning, different modalities have imbalanced and supplementary relation, with different information when drawing the same semantics. Learning explicit common subspace representations will easily overlook the modality-specific information. In this case, many works~\cite{peng2018modality, wu2020modality, yanagi2020enhancing, WangHKXP15, liu2018modality, xiong2020modality, PengQY18, NieWLHJY21, CaoLHH19, zhen2020deep, zhang2018collaborative, zhang2021label} construct the independent individual semantic space for each modality.

In particular, the work~\cite{peng2018modality} proposed an end-to-end framework that generates the modality-specific cross-modal similarity
without learning explicit common subspace and in each semantic subspace for modality, the model can preserve the modality-specific characteristics. $MS^{2} GAN$~\cite{wu2020modality} present a model that can learn modality-specific representation and modality-shared representation for each modality by training with a adversarial learning between generative and discriminative model. Generative model is to generate the semantic labels of representations, caption the inter-correlation and intra-correlation, and also differentiate the two different kinds of representations. Discriminative model is used to distinguish the modality of representations. The MsMFH~\cite{xiong2020modality} learns the modality-specific semantic representation for each modality and then applying correlation information to align the representations. Cross-modal hashing is a prevalent way to achieve cross-modal retrieval application, however, gaining the storage- and time-saving advantage may result in performance degradation because the finite-length hash code for similarity matrices cannot convey the correlation across modalities. The MLCAH model~\cite{ma2020multi} converts the multi-level correlation information to hash codes by learning global and local semantic information with multi-level information and label information. A single common space for learning optimal representation of visual and text features is not enough, this kind of methods will lead to loss key information of data, this work~\cite{yanagi2020enhancing} introduced a robust method to learn effective representations space for embeddings of visual and language by adopting two different types of generative models. Some existing methods learn implicit common space for cross-modal gap bridging that highly depend on the low-level features but overlook the internal characteristic in original data across modalities. This work~\cite{WangHKXP15} presents a new model to learn modality-specific representations by using two CNN structures to map the raw data. Hashing-based cross-modal retrieval methods also ignore the modality-specific features during bridging the heterogeneous gap. The MsPH~\cite{liu2018modality} model is able to maintain the local structure and internal correlation. To address the problem of manual labeling uncertainty, this model adopted a label enhancement mechanism. The heterogeneous gap comes from the inconsistent and complementary correlation across modalities, different amounts of the same level semantic information. Simply projecting them into a common space will lose their exclusive modality-specific information. This work~\cite{PengQY18} introduced a novel cross-modal similarity measurement by considering the modality-specific characteristics, which training through a end-to-end architecture to generate modalities-specific features, RNNs model and attention model are used to exploit the modality-specific for semantic space. Hashing code in cross-modal retrieval with modality-specific features will ignore the semantic relevance between two sets of modality-specific features. The DMFH model~\cite{NieWLHJY21} is a hashing based method that utilizes the multi-scale fusion models in each subnetwork to caption the semantic relevance. The DNN-based method~\cite{CaoLHH19} is a hybrid representation learning, which is used to learn shared space to generate embedding for each modality. This model employs stacked restricted Boltzmann machines to extract modality-specific features, a complicated network with joint autoencoders are applied, and the final representation for each modality is obtained by a stacked cross-modal autoencoder. To avoid the need for a large amount of annotation data for achieving expected performance, transferring the valuable information from effective annotation data to extend the dataset is considered. The DMTL~\cite{zhen2020deep} model transfers the previous labels to improve the retrieval system of new unseen labels by applying a pseudolabel algorithm.
The CSGH model~\cite{zhang2018collaborative}  is to learn the diversity of cross-modal features in the common spaces by capturing the individual modality-specific transformation. In detail, this model learns separate common subspaces for each modality by modality-specific transformation matrices then mapping these subspaces into a Hamming space with a common transformation matrix. In addition, the model captures the manifold structure with Laplacian regularization and the cross-modal relation with a correlation constraint. The unified hash codes from a shared space cannot fully obtain the characteristic of modality because of the internal heterogeneous difference. The LFMH model~\cite{zhang2021label} is a novel cross-modal hashing approach that learns modality-specific subspace by matrix factorization. Finally, the model can transform the cross-modal raw data into modality-specific features 

Different from the above methods, our work is inspired by the domain adaption network~\cite{bousmalis2016domain}, we learn two different common subspaces. To alleviate the loss of key information in only one common space learning, we learn additional implicit common space, where is for domain adaptation and weakly maintains the semantic representation.

% ~\cite{thomas2020preserving}

\section{APPROACH}
\label{approachsection}
\begin{figure}%
    \centering
    \includegraphics[width=\linewidth]{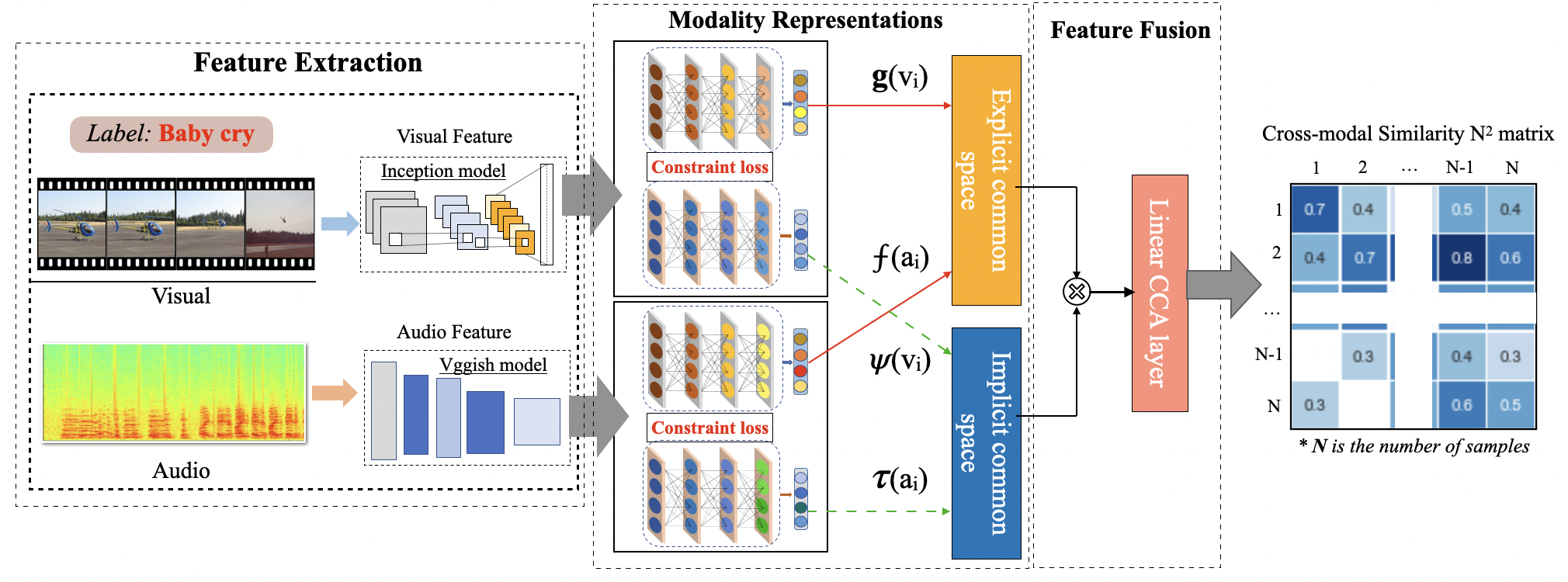}%
    \caption{The model consists of three parts: 1)Feature Extraction; 2)Modality Representations; 3)Feature Fusion. The model takes the global audio and visual representations extracted by Vggish and Inception model respectively, and projects each modality into two different latent subspaces: Explicit and Implicit common spaces. Then, these hidden representations are utilized to fuse features by linear CCA-layer to calculate cross-modal similarity.}%
    \label{fig:architecture}%
\end{figure}

\subsection{Problem Formulation}
Assume that there is a collection of $n$ videos, extracted audio and visual features from each video, denoted as $\Phi=\{\phi_{i}\}_{i=1}^{n}$, $\phi_{i}=(a_{i}, v_{i})$, where $a_{i}\in R^{d_{a}}$ is an audio feature and $v_{i}\in R^{d^{v}}$ is a visual feature, here $d_{a}$=128 and $d_{v}$=1024 are the dimension of input features. Each instance $\phi_{i}$ labeled by a semantic vector $Y_{i}=[y_{i1}, y_{i2}, ..., y_{ic}]$, where $c$ is the number of semantic categories. If the video $i$ belong to $y_{ij}$ $(1\leqslant j\leqslant c$), $y_{ij}$=1, otherwise $y_{ij}$=0. Finally, the above modalities and categories can be represented by feature matrix, $A=\{a_{i}, ..., a_{n}\} \in R^{d_{a}\times 128}, V=\{v_{i}, ..., v_{n}\} \in R^{d_{v}\times 1024}$.

Since the audio and visual representations $A$ and $V$ have inconsistent distributions and different dimensions, they are unable to compare with each other for the AVCMR task. It requires us to learn common space, explicit common space $S^{ex}$, and implicit common space $S^{im}$. In the first space, audio and visual representations are projected as $S_{ex}^{a}$ and $S_{ex}^{v}$ by $f(a_{i})$, $g(v_{i})$, respectively. In the second space, audio and visual representations are projected as $S_{im}^{a}$ and $S_{im}^{v}$ by $\psi(a_{i})$, $\tau(v_{i})$, respectively. Finally, the output of audio and visual representations is the fusion of $S_{ex}^{a}$\&$S_{im}^{a}$ and $S_{ex}^{v}$\&$S_{im}^{v}$, respectively.
\subsection{Proposed method}
The function of can be divided into three main parts: Feature Extraction (Section 4.3), Modality Representations and Modality Fusion (Section 3.3). The general framework of the is shown in Figure~\ref{fig:architecture}, and the configuration of is presented in Table~\ref{tab:config}.

\begin{table}[htp]
  \begin{center}
    \caption{Configuration of our proposed model.}
    \begin{tabular}{r|l} 
    \hline
       Log mel-spectrogram audio inputs& 96x64 \\
       Input of audio/visual branch& 128-D/1024-D \\
       Output of Ex-subspace& 10-D \\
       Output of Im-subspace& 10-D \\
       Hidden units for audio& [1024, 1024, 10] \\
       Hidden units for visual& [1024, 2048, 10] \\
       Output of CCA-layer& 10-D \\
      \hline
    \end{tabular}
    \label{tab:config}
  \end{center}
\end{table}
\subsection{Modality Representations and Fusion}
\textbf{Global-level Representations}. For audio and visual modality are extracted by pre-trained models (Section 4.3). The output of the features is frame-level features. Convert frame-level feature to global-level feature is as follows.
\begin{equation}
\begin{aligned}
  x_{g} = (m(x_{1}), m(x_{2}), ..., m(x_{l})),\\
  m(x_{i}) = \frac{1}{|d_{x}|}\sum_{j=1}^{d_{x}}(x_{ij})
   \label{eq:global}
\end{aligned}
\end{equation}
where $d_{x}$ is the dimension of $x_{i}$, $l$ is the number of frames.

\textbf{Explicit and Implicit Common Space.}
We now project global-level representations across modalities into two distinct common subspaces. Based on whether learning one-to-one pair correlation, we classify the common space into explicit- and implicit-subspaces. Firstly, the explicit common space is to learn the commonalities and correlation for each audio-visual pair extract from the same video. Secondly, the implicit common space captures the unique characteristics of that modality by minimizing the difference between label and feature.
\subsubsection{\textbf{\texorpdfstring{$L_{corr}$-Correlation loss}{Lg}}}. We apply correlation loss to reduces the cross-modal discrepancy, which makes the common audio-visual representation to be aligned close in the explicit common space. Technically, our correlation loss is CCA-variant loss as follows:

%$\sqrt{w^{T}_{x}\sum_{xx}w_{x}w^{T}_{y}\sum_{yy}w_{y}}$
\begin{eqnarray}
L_{corr} =  \frac{1}{n}\sum_{i=1}^{n}\frac{W^{T}_{a}\Sigma_{av}W_{v}}{\sqrt{W^{T}_{a}\Sigma_{aa}W_{a}\cdot W^{T}_{v}\Sigma_{vv}W_{v}}} 
\label{eq:corr}
\end{eqnarray}
where $W_{a}$, and $W_{v}$ are the transformation weight $\Sigma_{aa}$ and $\Sigma_{vv}$ are the covariance matrices of $A$ and $V$, respectively and $\Sigma_{av}$ is the cross covariance.

\subsubsection{\textbf{\texorpdfstring{$L_{dis}$-discriminative loss}{Lg}}}.
Preserve the discrimination of generated representation in the shared subspace is very important for AVCMR task. However, if we only learn the one-to-one pair correlation and overlook the category information during the training, the learned representation will far from our expectations. Here, we set a trick that is to make the dimension of output vector is the same as the label vector, such that the output vector is not only the generated representations but also it's the predicted label of the sample. We propose the following objective loss function to keep the discrimination of modality representations in the label space.
%$\psi(a_{i})$, $\tau(v_{i})$
\begin{eqnarray}
L_{dis} = \frac{1}{n}||\psi(a_{i})-y_{a_{i}}||_{F}+\frac{1}{n}||\tau(v_{i})-y_{v_{i}}||_{F}
\label{eq:dis}
\end{eqnarray}
where $||\cdot||_{f}$ represents the Frobenius-Norm.

\subsubsection{\textbf{\texorpdfstring{$L_{cons}$-constraint loss}{Lg}}}.
Loss $L_{corr}$ and $L_{dis}$ capture different aspects of the input. However, the end-to-end model training may cause redundancy of NN structures. In this case, we involve a constrain loss to enforce the two NN structures with the same input learned in the different direction followed by~\cite{bousmalis2016domain, liu2017adversarial, ruder2018strong}. Let $Q_{ex}^{m}$ and $Q_{im}^{m}$ be the matrix, we apply the orthogonality constraint for each modality pair as:
\begin{eqnarray}
||(Q_{ex}^{m})^{T} Q_{im}^{m}||_{F}^{2}
\label{eq:consa}
\end{eqnarray}
where $||\cdot||_{f}^{2}$ is the squared Frobenius-Norm. In addition, we add the constraint in the implicit common space between audio and visual representation learning. The final constraint loss is calculated as:
\begin{eqnarray}
L_{cons} = ||(Q_{ex}^{m})^{T} Q_{im}^{m}||_{F}^{2} + ||(Q_{im}^{m1})^{T} Q_{im}^{m2}||_{F}^{2}
\label{eq:cons}
\end{eqnarray}

Incorporating above 3 loss equation ~(\ref{eq:corr}, ~\ref{eq:dis} and ~\ref{eq:cons} and assign the weight for each loss based on the output loss value. The overall loss function of our model is as:
\begin{eqnarray}
L=L_{corr}+\alpha L_{dis}+\beta L_{cons}
\label{eq:overall}
\end{eqnarray}
Here, $\alpha$, $\beta$ are the interaction weights that decide the contribution of each component to the loss $L$, we will discuss these hyperparameter in Section 5.2.4. Finally, the final loss function $L$ can be optimized using a SGD algorithm.

\textbf{Feature fusion.} Once we get the output representation from two subspaces, it requires us to fuse them into a joint embedding. We apply a simple fusion mechanism by adding a linear-CCA layer and the input of the layers is the concatenation of representations from two subspaces for each modality. Our system will output 10-D representation for each modality for cross-modal similar calculation.

\section{EXPERIMENTS}
\label{experimentsection}
\subsection{Datasets}
\begin{figure}%
    \centering
    \subfloat[\centering ]{{\includegraphics[width=6cm]{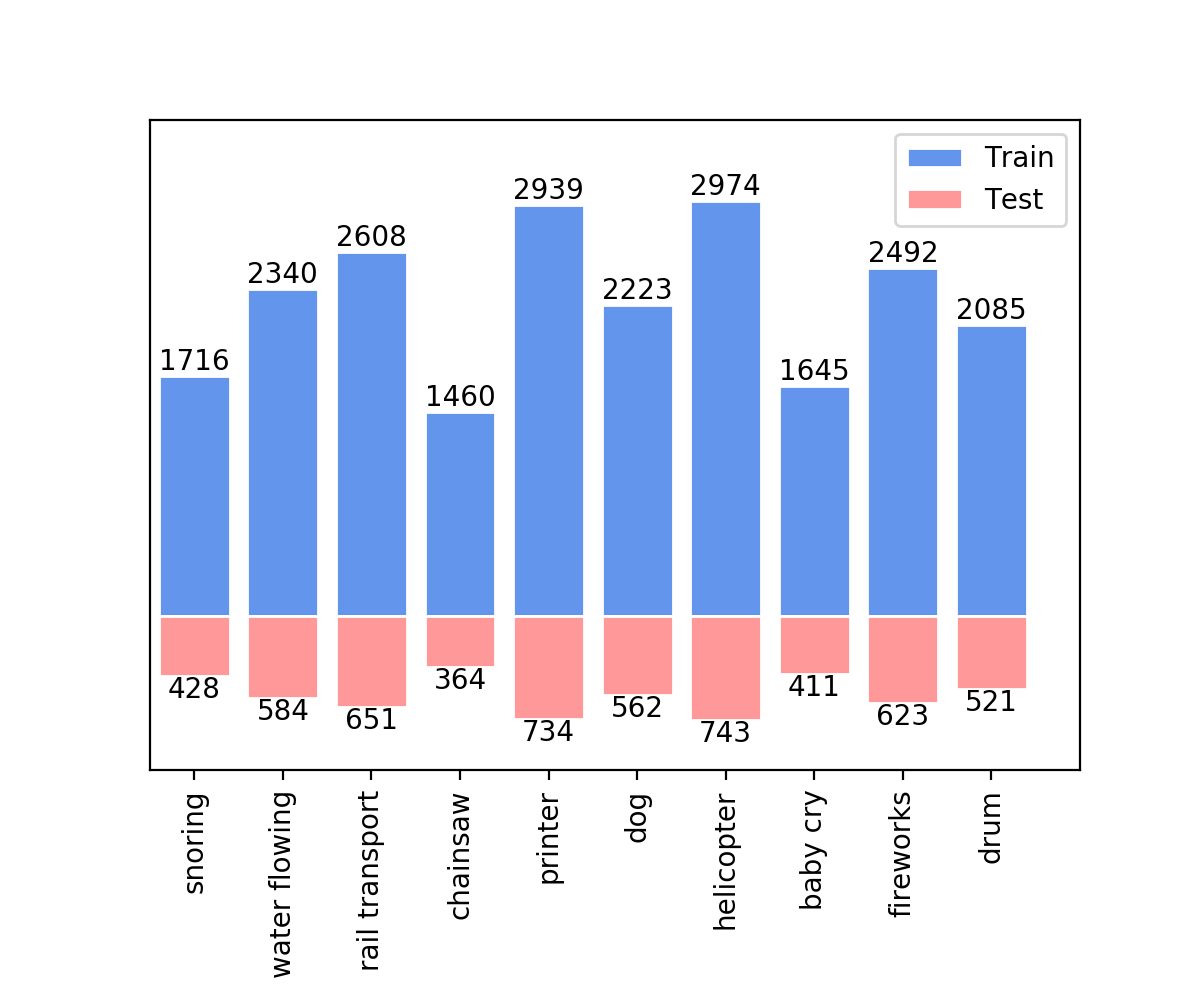} }}%
    \subfloat[\centering ]{{\includegraphics[width=6cm]{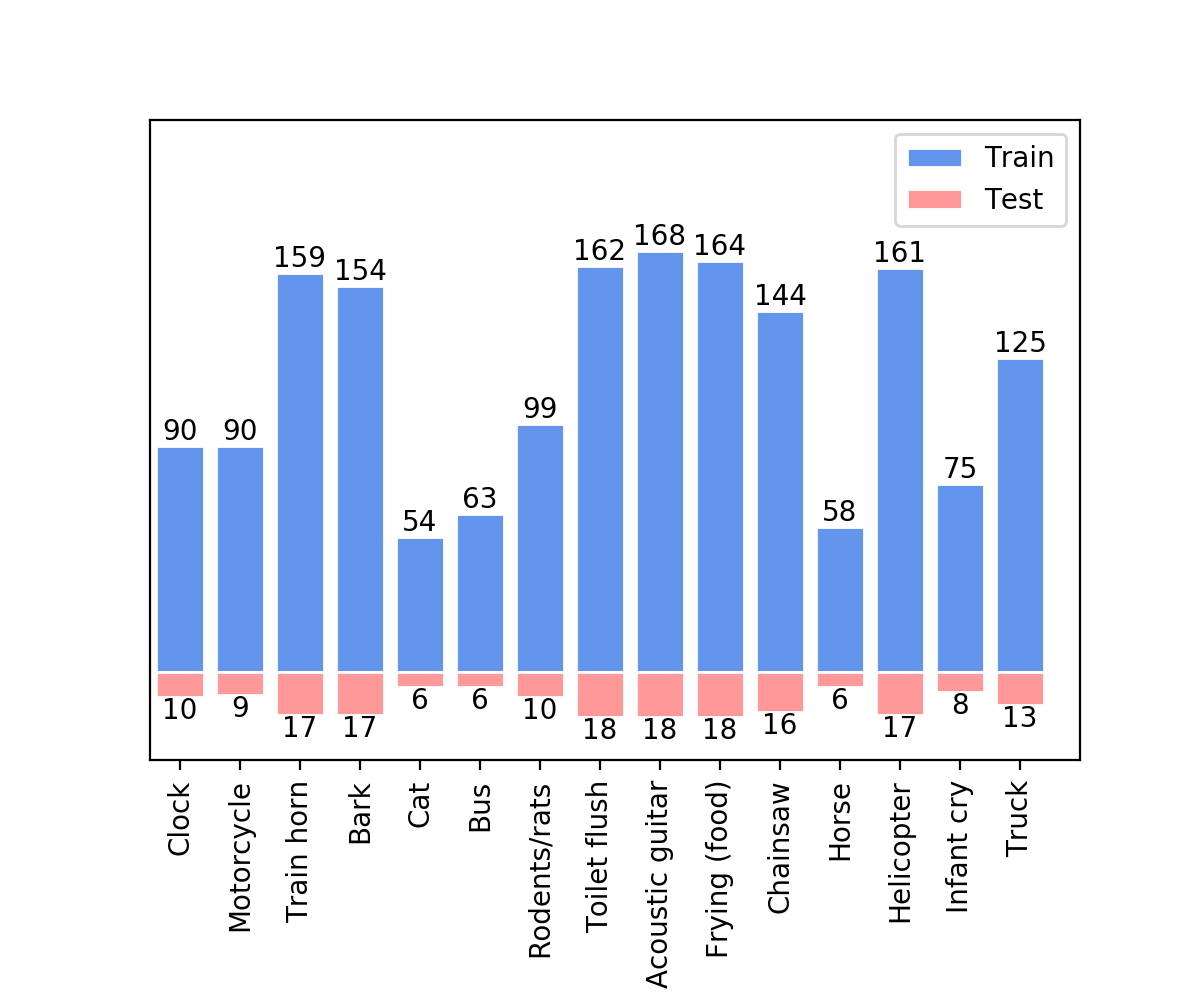} }}%
    \caption{The number of videos with different categories in train set and testing set. (a) VEGAS dataset, (b) AVE dataset.}%
    \label{fig:dataset}%
\end{figure}
We evaluate our model on two different video datasets: VEGAS~\cite{zhou2018visual} dataset and AVE-dataset~\cite{tian2018audio}. As for VEGAS dataset, we follow the work~\cite{zeng2020deep} to extract each video with audio feature and visual feature by using advanced pre-trained models and then apply the global feature by frame-axis average as the final representations, more details about the two audio-visual dataset are as follows.
\subsubsection{\textbf{VEGAS dataset}}
The VEGAS dataset is a subset of Google Audioset and employ Amazon Mechanical Turks to clean the data. This dataset encompasses 10 human voices vehicle sounds, and nature sounds categories (chainsaw, rail transport, helicopter, baby cry or infant cry, drum, snoring, printer, fireworks, water flowing, and dog.) that are labeled by audio and visual double tracks check. Moreover, The length of a video ranges from 2 to 10 seconds and the average length overall is around 7 seconds. This dataset contains 28,103 videos and each video is labeled by a single label. We divided all the videos into train and test set by 80\% and 20\% and keep each category is balance appeared in the train and test sets, seen in Figure~\ref{fig:dataset}(a). Finally, we apply 5,621 videos for testing and 22,482 videos for training in our experiments.
\subsubsection{\textbf{AVE dataset}}
Audio-Visual Event (AVE) dataset contains 4,143 videos covering 28 event categories, each video is temporally annotated with audio-visual event boundaries. The lengths of video are range from 2 seconds to 10 seconds. In our experiment, we remove the music-related categories and keep 15 categories (Clock, motorcycle, train horn, bark, cat, bus, rodents/rats, toilet flush, acoustic guitar, frying (food), chainsaw, horse, helicopter, infant cry, truck). We select 189 videos as testing set, 1766 videos as train set, and keep category balance in each set, seen in Figure~\ref{fig:dataset}(b).

\subsection{Evaluation Metric}
To evaluate our proposed model, we apply Mean Average Precision (MAP) as the main computational metric for audio-visual cross-modal retrieval on both VEGAS dataset and AVE dataset. In addition, to further leverage the  performance of our model, we follow the work~\cite{zeng2020deep} to take Time Cost as another metric to evaluate real-time started from training model to test the model on the VEGAS dataset. Besides, we explore Precision-Scope Curves (PSC) and Average Precision (AP) per category to leverage our architecture in order to further investigate our proposed model. During the testing, the system produces a rank list in one modality for each query in another modality, we regard the candidates in the rank list with the same category as query is correct.
\subsection{Feature Extraction}
To make a fair comparison, we apply the same audio-visual features as input of all models. Because all the compared methods are required learning common spaces, so the input audio-visual representations are global features.
\subsubsection{\textbf{Audio Features}}
We utilize the \footnote{https://librosa.org/}{Librosa Library} to extract Mel spectrogram feature as the input of Vggish model~\cite{hershey2017cnn}, with default parameters: hop size=512, nftt=2,048 and etc. The pre-trained VGG-like model is trained on AudioSet benchmark for audio classification task. Finally, audio representation is a 128-D global feature.

\subsubsection{\textbf{Visual Features}}
We extract frame-level visual features by using the public Inception V3~\footnote{https://github.com/fchollet/deep-learning-models} model~\cite{abu2016youtube}. which is trained on the ImageNet dataset. Here, we decode each video at one frame-per-second, so the number of frames is the equal to the number of seconds, then we feed the features into Inception V3 network and use ReLu activation in the last hidden layer. The frame-level features of Inception V3 model is 2048-D per frame and discard the motion information. Afterwards, we exploit Principal component analysis (PCA) to reduce the dimension to 1024, so the final frame-level visual features is 1024*$len$, where $len$ is the number of frames. In our experiment, we apply global 1024-D feature obtained by frame-axis average as visual representations.

\subsection{Baselines}
We comprehensively compare our model with various baselines, including Random case, CCA-variant methods and other state-of-the-art methods, the detail is as follows. As for the Random case, instead of getting the rank list for each query by the cross-modal similarity, it is by random.

\subsubsection{\textbf{CCA-variant Models}}
Multitudinous models have been proposed for cross-modal learning, CCA-variant as traditional methods for learning the correlation in two variables across modalities should be taken into account, so that to conduct a comprehensive comparative study against our model. Prevalent CCA-variant baselines include:
\begin{itemize}
  \item \textbf{CCA~\cite{hardoon2004canonical}} (Canonical Correlation Analysis) is to find basis vectors for two variable sets to ensure the correlation between the projections of the two variables on basis vectors are mutually maximised. An alternative description is that CCA is to find linear transformations for two multivariate sets to optimize the correlation between them, which is the unsupervised learning and strongly associate to mutual information.
  
  \item \textbf{KCCA~\cite{akaho2006kernel}} (Kernel Canonical Correlation Analysis) differ from CCA in finding the non-linear correlation by the kernel method, such as the known method ”kernel trick”, which is to improve the computational power of the linear CCA and extract more useful information of the input data by projecting the data into a high-dimensional space.
  
  \item \textbf{DCCA~\cite{andrew2013deep}} (Deep Canonical Correlation Analysis) is to obtain the nonlinear transformations of two data sets by deep learning and CCA-linear layer, which can be regard as a nonlinear extension of the linear CCA, different from CCA, DCCA is a parametric model and did not require the inner product. In addition, DCCA can be loosely viewed as learning a kernel for KCCA and offer a flexible nonlinear alternative to KCCA.
  
  \item \textbf{C-CCA~\cite{rasiwasia2014cluster}} (Cluster Canonical Correlation Analysis) is to cluster data across modalities into several classes and tries to enhance the intra-cluster correlation. Unlike the pairwise learning between data points from difference modalities, C-CCA is to divide the cross-modal data points into several clusters and learn new representations to maximize the correlation between two data modalities while set the data points from different classes apart in the common space.
  
  \item \textbf{C-DCCA~\cite{yu2018category}} (Cluster Deep Canonical Correlation Analysis)is a combination of Cluster-CCA and DCCA, which segregate the non-linear representation from deep learning method into several related clusters to optimize the correlation, where cross-modal data are non-linear projected into a common space while the data points from the same cluster have high correlation.
  
  \item \textbf{TNN-C-CCA~\cite{zeng2020deep}} (Triplet Neural Network \& Cluster Canonical Correlation Analysis) can be viewed as an extension of C-CCA and C-DCCA, which considers the category information from both similar and dissimilar correlation. TNN-C-CCA presents an audio-visual specific loss function to improve the Cluster-CCA and is the present state-of-the-art method on VEGAS dataset, where the output of cluster CCA are mapped into a latent shared subspace through a triplet neural network while the data points from different clusters are highly correlated and the data points from different clusters are little correlated.
\end{itemize}
\subsubsection{\textbf{Advanced methods}}
Numerous models of cross-modal learning are presented in recent years, we list some advanced models that publish in multimedia domain of top conferences as baselines.
\begin{itemize}
  \item \textbf{UGACH~\cite{ZhangPY18}} (Unsupervised Generative Adversarial Cross-modal Hashing approach) utilizes Generative Adversarial Network to exploit the potential manifold structure of data across modalities by unsupervised learning. Given a data point from on modality, the generative mechanism will fit the manifold structure of the distribution and select the correlated data points from another modality to feint the discriminative model.
  \item \textbf{AGAH~\cite{GuGGLXW19}} (Adversary Guided Asymmetric Hashing) applies a multi-label attention model with adversarial learning to enhance discrimination of cross-modal representations, which can generate hashing codes to fully keep the multi-label information of all the data points.
  
  \item \textbf{UCAL~\cite{XLYSS17} and ACMR~\cite{wang2017adversarial}} (Unsupervised cross-modal retrieval and Adversarial Cross-Modal Retrieval) apply unsupervised and supervised adversarial learning for cross-modal representation learning. A feature projector of ACMR to learn modality-invariant representation while modality classifier try to distinguish the modality of generated representation. 
  
  \item \textbf{DSCMR~\cite{zhen2019deep}} (Deep Supervised Cross-modal Retrieval) is advanced in learning discriminative features in label space and feature common space by supervised learning that minimizes the discrimination loss in representation learning space and label learning space while bridge the modality gap by using a weight sharing strategy. In our experiment, DSCMR model achieves the best performance on AVE dataset.
  
\end{itemize}

\section{RESULTS AND ANALYSIS}
\label{results&analysis}
\subsection{Comparison with Existing Methods}
The comparative results for AVCMR task on VEGAS dataset and AVE dataset are presented in Table~\ref{tab:Results}, it shows that our model achieves the best performances and significantly outperforms all CCA-variant methods including the state-of-the-art method TNN-C-CCA, and others advanced methods on both audio-visual datasets. In particular, our model gains 4.6\% of MAP in $audio\rightarrow visual$ retrieval, 4.1\% of MAP in $visual\rightarrow audio$ retrieval, 4.3\% of MAP in average MAP, over TNN-C-CCA on VEGAS dataset. Also, our model outperforms the GOAT model DSCMR on AVE dataset by the same 2.3\% of MAP in $audio\rightarrow visual$ and $visual\rightarrow audio$ retrieval.

\begin{figure}%
    \centering
    \subfloat[\centering ]{{\includegraphics[width=7cm]{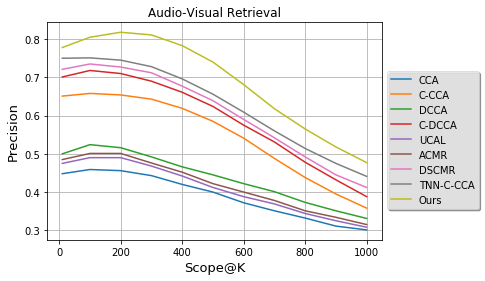}}}%
    \subfloat[\centering ]{{\includegraphics[width=5.5cm]{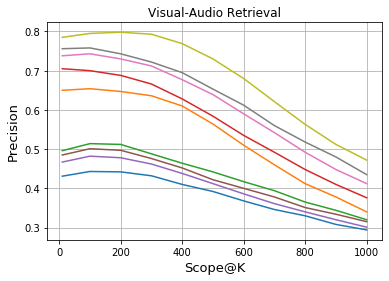}}}%
    \caption{Precision-scope curves on the VEGAS dataset for (a) audio2visual and visual2audio retrieval experiments, with K ranges from 10 to 1000.}%
    \label{fig:PSC}%
\end{figure}

Except for the evaluation with MAP, we apply precision-scope curves as an additional assessment. Figure~\ref{fig:PSC} shows the curves of and of CCA, C-CCA, DCCA, C-DCCA, UCAL, ACMR, DSCMR, TNN-C-CCA. The precision-scope curves and the MAP are consistent when audio or visual as query and our model have significant gains.

The supervised model is not completely learning representation with label in each common space. In the explicit common space, our model projects modalities into this subspace to optimize the correlation of pair to pair. In Table ~\ref{tab:baselines}, the variant model only with explicit common space is unsupervised learning. Compared with the other unsupervised learning methods CCA, KCCA, UGACH and UCAL, variant model also can achieve the best performance.

\begin{table}
  \caption{The MAP of Audio-Visual Cross-Modal retrieval for Our Proposed Model and Some Previous State-of-the-art Models on VEGAS Dataset and AVE Dataset}
  \begin{tabular}{c|ccc|ccl}
    \toprule
   \multirow{2}{*}{\textbf{Models}} & \multicolumn{3}{|c}{VEGAS Dataset} & \multicolumn{3}{|c}{AVE Dataset} \\
    
    \cline{2-7}
%   & audio$\rightarrow$visual
%   & visual$\rightarrow$audio
%   & Average
%   & audio$\rightarrow$visual
%   & visual$\rightarrow$audio
%   & Average
  & A$\rightarrow$V
  & V$\rightarrow$A
  & Avg.
  & A$\rightarrow$V
  & V$\rightarrow$A
  & Avg.
   \tabularnewline 
\hline

    % \midrule
    \texttt{Random} 
    & 0.110 & 0.109 & 0.109     
    & 0.127 & 0.124 & 0.126\\
    \texttt{CCA~\cite{hardoon2004canonical}}    
    & 0.332 & 0.327  & 0.330    
    & 0.190 & 0.189 & 0.190 \\
    \texttt{KCCA~\cite{akaho2006kernel} }   
    & 0.288 & 0.273  & 0.281    
    & 0.133 & 0.135 & 0.134 \\
    \texttt{DCCA~\cite{andrew2013deep} }   
    & 0.478 & 0.457  & 0.468   
    & 0.221 & 0.223 & 0.222 \\
    \texttt{C-CCA~\cite{rasiwasia2014cluster}}  
    & 0.711 & 0.704  & 0.708    
    & 0.228 & 0.226 & 0.227\\
    %& 0.652 & 0.654  & 0.653    
    %& 0.204 & 0.198 & 0.201\\
    \texttt{C-DCCA~\cite{zeng2018audio}} 
    & 0.722 & 0.716  & 0.719    
    & 0.230 & 0.227 & 0.229\\
    \hline
    
    \texttt{UGACH~\cite{ZhangPY18}}  
    & 0.182 & 0.179 & 0.181     %VEGAS dataset 
    & 0.165 & 0.159 & 0.162\\   %AVE dataset
    \texttt{AGAH~\cite{GuGGLXW19}}   
    & 0.578 & 0.568 & 0.573      
    & 0.200 & 0.196 & 0.198\\
    \texttt{UCAL~\cite{XLYSS17}}   
    & 0.446 & 0.436 & 0.441      
    & 0.153   & 0.150   & 0.152\\
    \texttt{ACMR~\cite{wang2017adversarial}}   
    & 0.465 & 0.442 & 0.454      
    &  0.162   & 0.159   & 0.161\\
    \texttt{DSCMR~\cite{zhen2019deep}}   
    & 0.732 & 0.721 & 0.727      
    & 0.314 & 0.256 & 0.285\\
    \texttt{TNN-C-CCA~\cite{zeng2020deep}} 
    & 0.751 & 0.738 & 0.745      
    & 0.253 & 0.258 & 0.256\\
    
    \hline
    \texttt{\textbf{Our model}}   
    & \textbf{0.797} & \textbf{0.779} & \textbf{0.788} 
    & \textbf{0.337} & \textbf{0.279} & \textbf{0.308}\\
    \bottomrule
  \end{tabular}
  \label{tab:Results}
\end{table}
%%%%%%%%%%%%%%%%%%%%%%%%%%%%%%%%%%%%
%%%%%%%%%%%%%%%%%%%%%%%%%%%%%%
\subsection{Ablation Study}
\subsubsection{\textbf{Role of Time Cost}}
Table~\ref{tab:config} shows the fused representation of our model is set as 10 and the same as other models. To fair comparison of time cost, we only consider the time consuming during the model training. Table~\ref{tab:timecost} presents the Time-consuming (minutes) of model training on VEGAS dataset. Our model only takes around 8 minutes to achieve the model training, which only falls behind the C-CCA and DCCA. Our model saves time significant over state-of-the-art TNN-C-CCA that takes 192.0 minutes to train the final model.
\begin{table}
  \caption{Time Cost of our model and Some Baselines model training on VEGAS dataset.}
  \begin{tabular}{cc|cl}
    % \toprule
    Models&Time(Minutes)&Models&Time(Minutes)\\
    \midrule
    Random         & 0.04  & CCA            & 1.08  \\
    DCCA           & 5.15  & KCCA           & 2.03 \\
    ACMR           & 176.7 & TNN-C-CCA      & 192.00 \\
    C-CCA          & 1.24  & C-DCCA         & 18.56 \\
    DSCMR          & 22.05 & Explicit-space & 6.15 \\
    Implicit-space & 6.77  & Our model      & 8.36 \\
  \bottomrule
\end{tabular}
\label{tab:timecost}
\end{table}
\subsubsection{\textbf{Role of Common Space}}
The two subspaces of our model are achieved by the three-terms combined objective function. Each term is consistent with the function of each subspace. To further investigate our model, we design three variations of our model. Our model-invariant (Explicit-space and Implicit-space) share the same neural network (NN) structures with our model and apply the combinations (correlation loss with constraint loss, discriminative loss with constraint loss) respectively. invariant model (Ex-Im-share) means the same modality shares the same NN structures for two subspaces and utilizes the same loss function with our model. From the results in Table~\ref{tab:baselines}, our model combines the three terms will stimulate the advantage of representation learning in two common subspaces and improve the performance of AVCMR task.

\begin{figure}%
    \centering
    \subfloat[\centering ]{{\includegraphics[width=10cm]{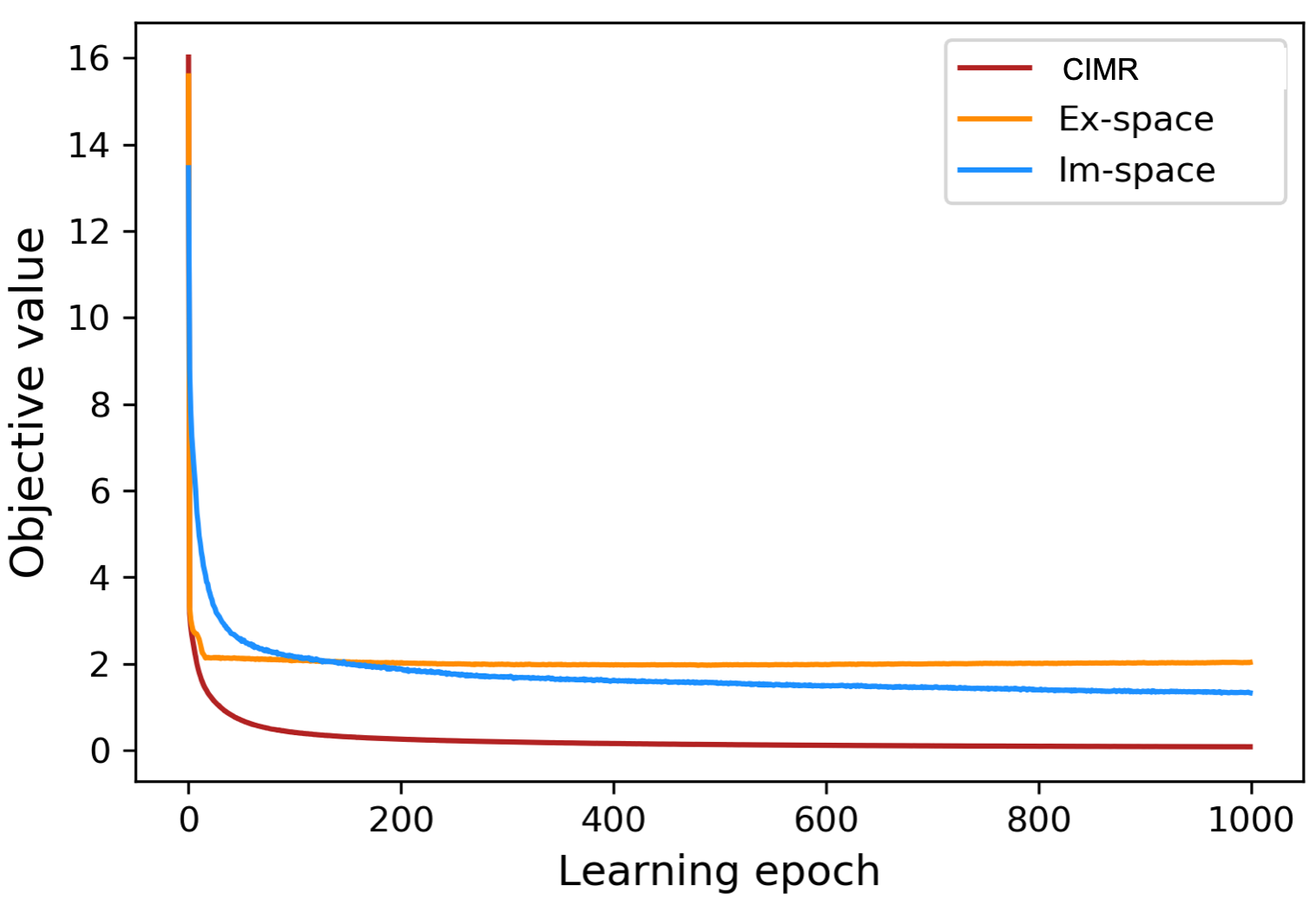} }}%
    \caption{The objective function value of our model with the change of training epochs [1, 1000], on the VEGAS dataset.}%
    \label{fig:loss_convergence}%
\end{figure}

Our model jointly optimize the correlation loss and discriminative loss in the objective function, To further exploit the effect of multi-subspaces in our model, we sample the correlation loss and the discriminative loss values from epoch ranges from 1 to 1000, seen in Figure~\ref{fig:loss_convergence}. The loss and other two loss combination decrease monotonously and smoothly converge. The loss curves are consistent with the expectation of two subspaces combination. 

\begin{table}
  \caption{The MAP of Audio-Visual Cross-Modal retrieval for Our Proposed Model and Some Baselines on VEGAS Dataset and AVE Dataset}

  \begin{tabular}{c|ccc|ccl}
    \toprule
   \multirow{2}{*}{\textbf{Models}} & \multicolumn{3}{|c}{VEGAS Dataset} & \multicolumn{3}{|c}{AVE Dataset} \\
    
    \cline{2-7}
  & A$\rightarrow$V
  & V$\rightarrow$A
  & Avg.
  & A$\rightarrow$V
  & V$\rightarrow$A
  & Avg.
   \tabularnewline 
\hline

    % \midrule
    \texttt{Random} & 0.110 & 0.109 & 0.109             & 0.127 & 0.124 & 0.126\\
    \texttt{Explicit-space} & 0.687 & 0.692 & 0.690     & 0.232 & 0.244 & 0.238\\
    \texttt{Implicit-space} & 0.614 & 0.622 & 0.618     & 0.222 & 0.263 & 0.242\\
    \texttt{Ex-Im-share} & 0.781 & 0.773 & 0.777 & 0.278 & 0.243 & 0.261\\
    \texttt{Our model}  & \textbf{0.797} & \textbf{0.779} & \textbf{0.788} & \textbf{0.337} & \textbf{0.279} & \textbf{0.308}\\
    \bottomrule
  \end{tabular}
    \label{tab:baselines}
\end{table}
\subsubsection{\textbf{Role of Category}}
\begin{figure}%
    \centering
    \subfloat[\centering ]{{\includegraphics[width=0.5\textwidth]{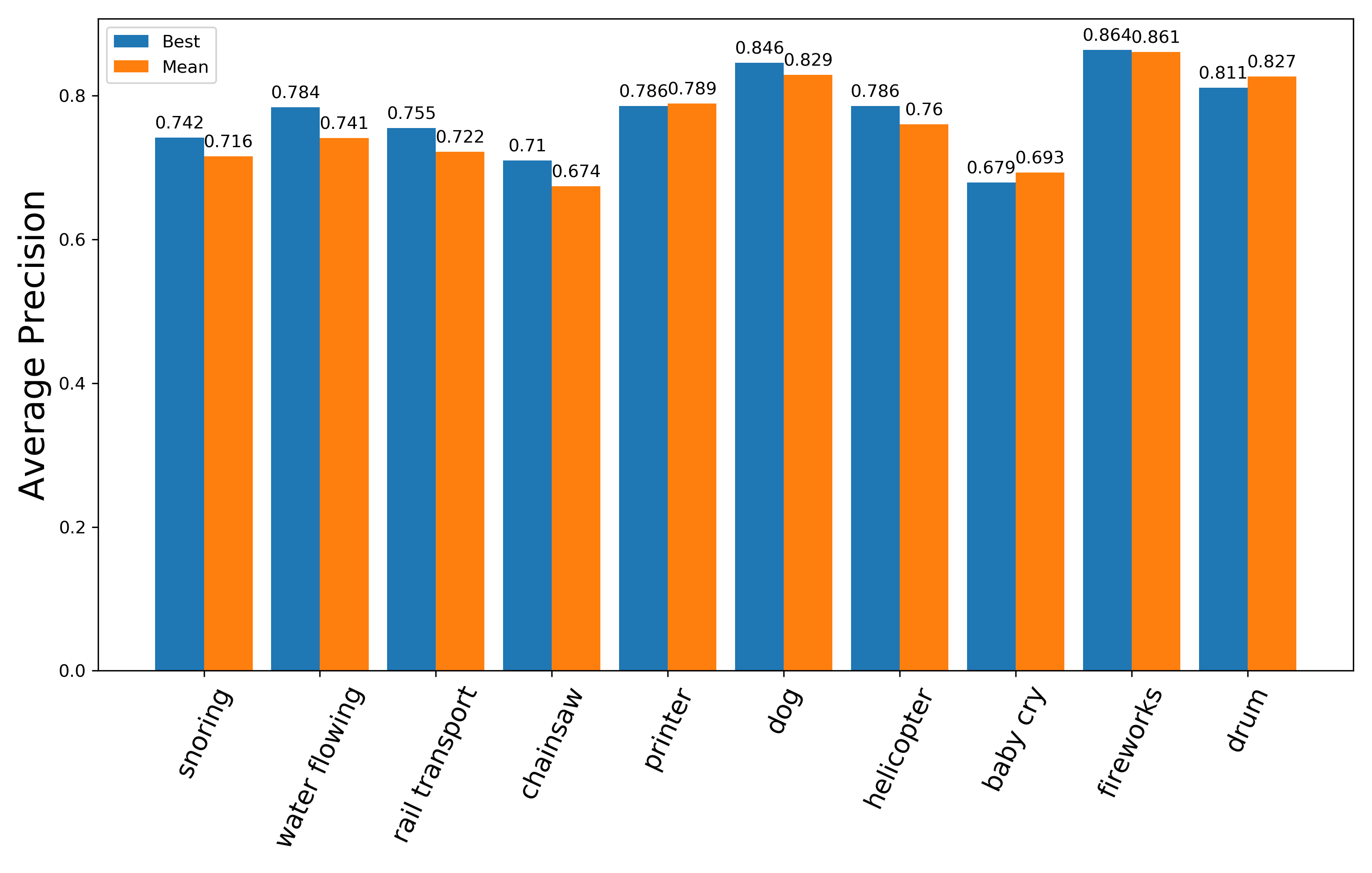}}}%
    \subfloat[\centering ]{{\includegraphics[width=0.5\textwidth]{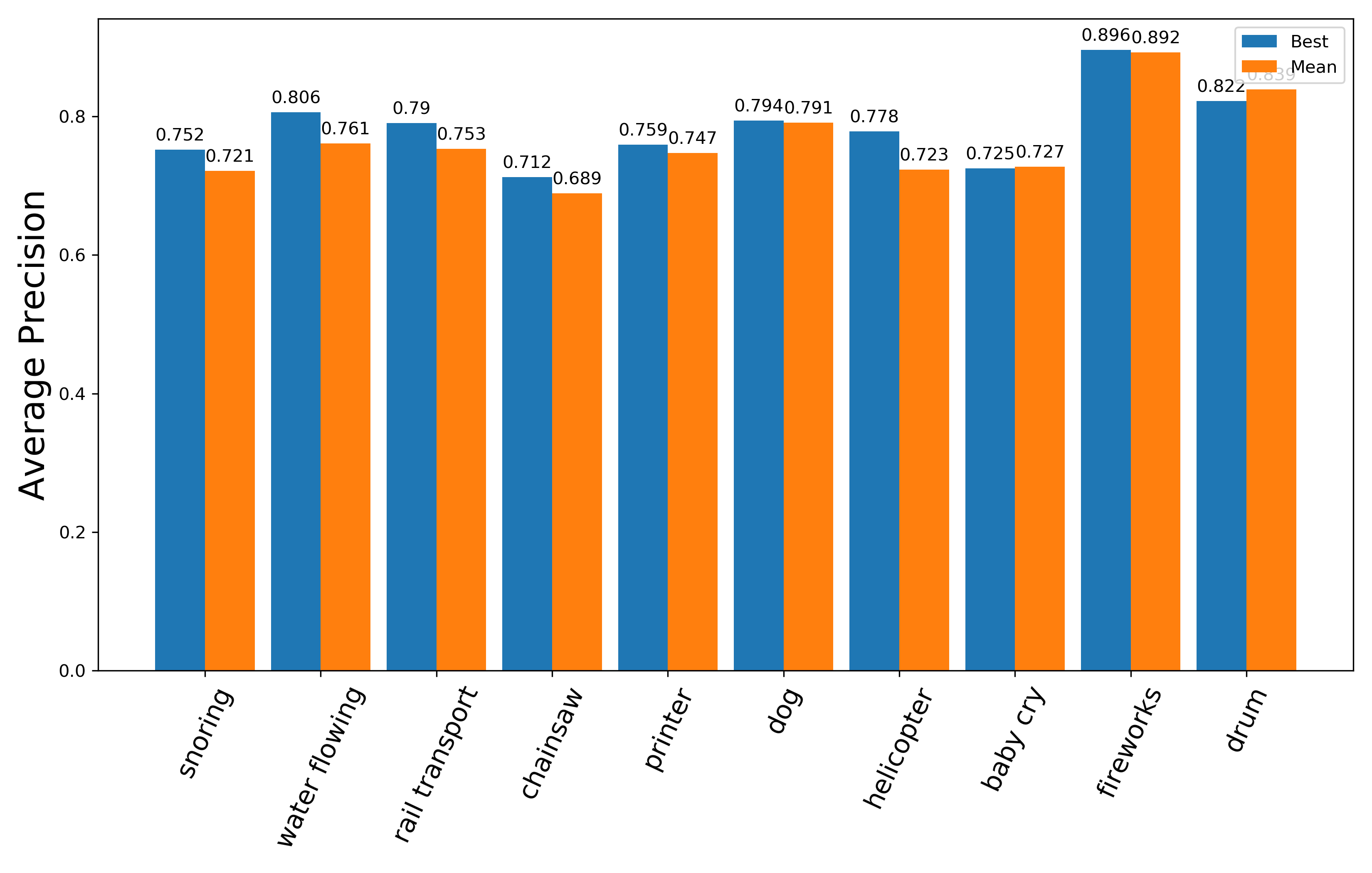}}}%
    \caption{Summary of the Audio-Visual Retrieval Results by Class. For each class two values are shown: the maximum AP obtained by our model(max) and the Median AP over all Our Baselines (mean).}%
    \label{fig:categoryx}%
\end{figure}

In Figure~\ref{fig:categoryx}, we show the average precision (AP) per category of our model and its variant models on VEGAS dataset. In particular, the Best AP perform by our model, the Mean AP is the mean of our model and three variant mentions in subsection 5.2.2. We divide the dataset into train and testing set and keep the category balance, the figure can roughly show the higher AP indicates the easier to retrieve. We can see the Dog barking and the firework get the highest AP with our model and the overall variant methods when audio as query. Take look at the Dog barking, audio2visual can get 84.6\% with our model but visual2audio only 79.4\%, almost 5\% difference.

\subsubsection{\textbf{Role of Model Parameters.}} 
\begin{figure}
     \centering
     \begin{subfigure}[b]{0.3\textwidth}
         \centering
         \includegraphics[width=\textwidth]{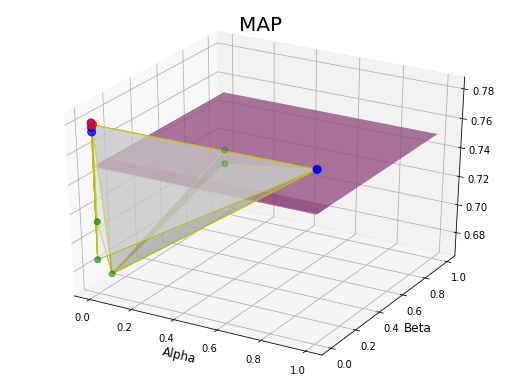}
         \caption{Audio-Visual (VEGAS)}
         \label{fig:y equals x}
     \end{subfigure}
     \hfill
     \begin{subfigure}[b]{0.3\textwidth}
         \centering
         \includegraphics[width=\textwidth]{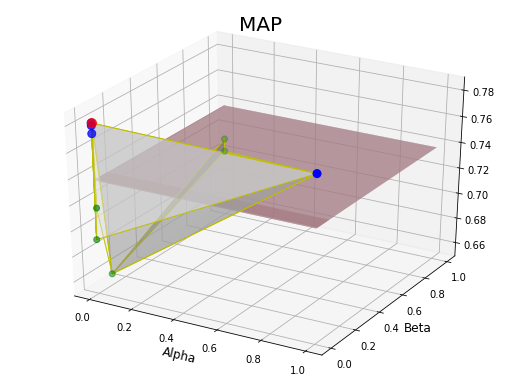}
         \caption{Visual-Audio (VEGAS)}
         \label{fig:three sin x3}
     \end{subfigure}
     \hfill
     \begin{subfigure}[b]{0.3\textwidth}
         \centering
         \includegraphics[width=\textwidth]{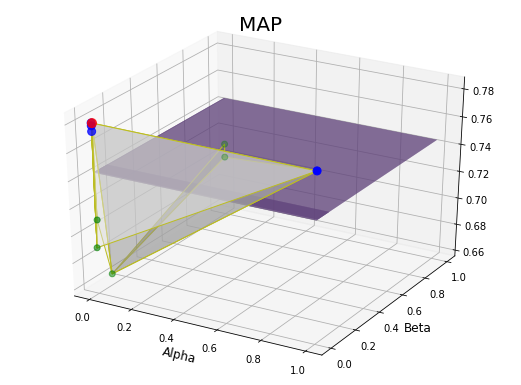}
         \caption{Average (VEGAS)}
         \label{fig:five over x1}
     \end{subfigure}
     \medskip
     \begin{subfigure}[b]{0.3\textwidth}
         \centering
         \includegraphics[width=\textwidth]{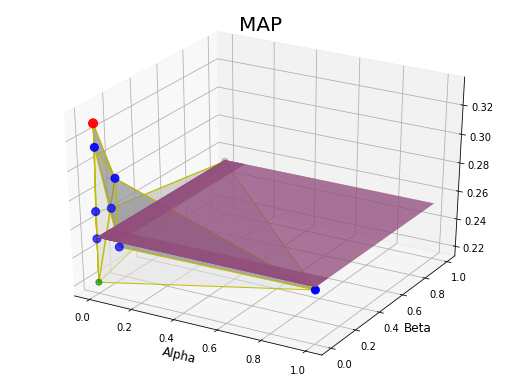}
         \caption{Audio-Visual (AVE)}
         \label{fig:five over x2}
     \end{subfigure}
     \begin{subfigure}[b]{0.3\textwidth}
         \centering
         \includegraphics[width=\textwidth]{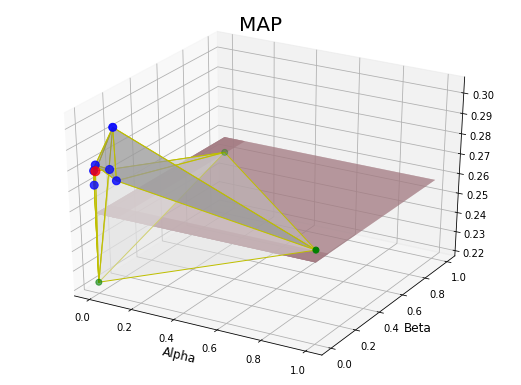}
         \caption{Visual-Audio (AVE)}
         \label{fig:five over x3}
     \end{subfigure}
     \begin{subfigure}[b]{0.3\textwidth}
         \centering
         \includegraphics[width=\textwidth]{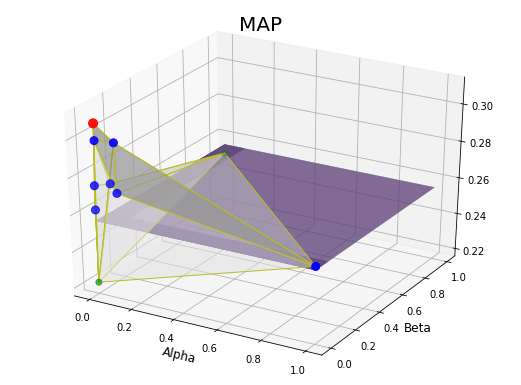}
         \caption{Average (AVE)}
         \label{fig:five over x4}
     \end{subfigure}
        \caption{The MAP Figures of with the Changing of Parameters: Alpha and Beta. The Planes Represent the MAP of the State-of-the-art model TNN-C-CCA, the Green Circles Denote the Situation (Fixed Alpha and Beta) that the MAP of our Model is lower than TNN-C-CCA Model, the Blue Circles Show the Higher and the Red Circles Imply the Best Case.}
        \label{fig:ab}
\end{figure}

Alpha ($\alpha$) and Beta ($\beta$) is a couple of hyper-parameters in loss function of our model. As $\alpha$ and $\beta$ distribute the contribution of the difference between two subspace and the modality-specific characteristic remains, we try to analyze the effect of them on cross-modal representation generation during training on two datasets (VEGAS and AVE). Here, we sample $\alpha$, $\beta$ from {0, 0.001, 0.01, 0.1, 1}, note that $\alpha$=0, our model is the (explicit-space) model. The figure shows that when $\alpha$, $\beta$ are set as 0.1, 0.01 or 0.001 can surpass the state-of-the-art TNN-C-CCA model, within the couple, when $\alpha$=0.01, $\beta$=0.001, our model can get the best performance on both datasets, seen in Figure~\ref{fig:ab}.

\begin{figure}
     \centering
     \begin{subfigure}[t]{0.3\textwidth}
         \centering
         \includegraphics[width=\textwidth]{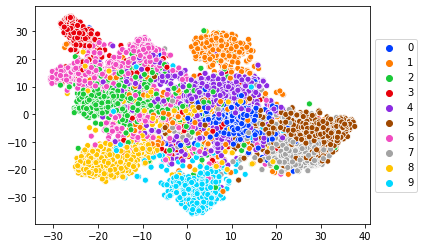}
         \caption{Original Audio Samples}
         \label{fig:y equals x1}
     \end{subfigure}
     \hfill
     \begin{subfigure}[t]{0.3\textwidth}
         \centering
         \includegraphics[width=\textwidth]{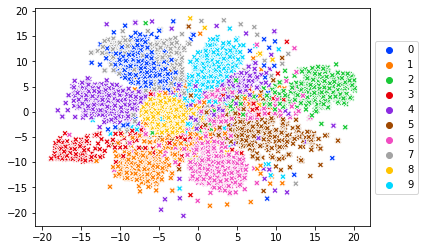}
         \caption{Original Visual Samples}
         \label{fig:three sin x1}
     \end{subfigure}
     \hfill
     \begin{subfigure}[t]{0.3\textwidth}
         \centering
         \includegraphics[width=\textwidth]{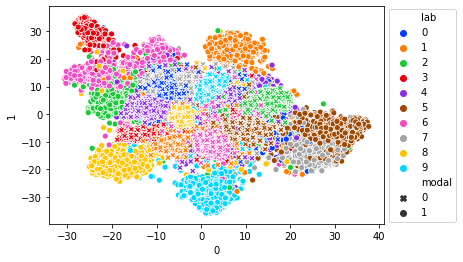}
         \caption{Original Audio-Visual Samples}
         \label{fig:three sin x2}
     \end{subfigure}
     \medskip
     
     \begin{subfigure}[t]{0.3\textwidth}
         \centering
         \includegraphics[width=\textwidth]{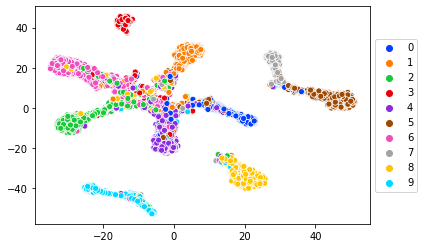}
         \caption{Audio Representations}
         \label{fig:five over x01}
     \end{subfigure}
     \hfill
     \begin{subfigure}[t]{0.3\textwidth}
         \centering
         \includegraphics[width=\textwidth]{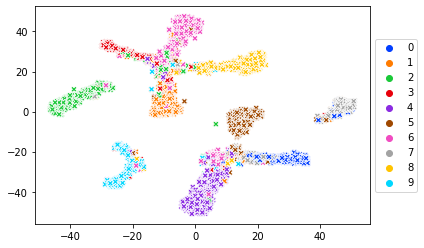}
         \caption{Visual Representations}
         \label{fig:five over x02}
     \end{subfigure}
     \hfill
     \begin{subfigure}[t]{0.3\textwidth}
         \centering
         \includegraphics[width=\textwidth]{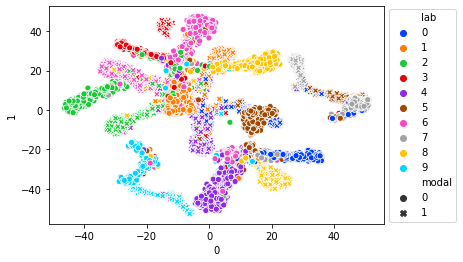}
         \caption{Audio-Visual Representations}
         \label{fig:five over x03}
     \end{subfigure}
     
        \caption{t-SNE visualization for the test data in the VEGAS Dataset. It includes original audio/visual sample and audio/visual representations.}
        \label{fig:final}
\end{figure}

\subsubsection{\textbf{Visualisation of the Learned Representation}}
To investigate the effectiveness of representations fusion from two subspaces of our model, we visualize the change of representations distribution from testing set during our model training on the VEGAS dataset by t-SNE tool. A comparison between the first row and second row in Figure~\ref{fig:final}, which indicates our model separate data points into semantic clusters and some of them are segregated from other clusters with clear boundary, but there are still some clusters will intersect with each other and hard to segregate. The segregated cluster will get easier to retrieve the related documents and the MAP will be higher than those intersected clusters. The AVCMR in this work is category-based retrieval, the ideal generated representation of the proposed model is all the clusters are segregated and the distance of boundary is enough for the system to distinguish the relevant candidates.

\section{CONCLUSION}
\label{conclusion}
In this paper, we presented a comprehensive audio-visual representation learning model that projects the modalities into explicit common space and implicit common space, the learned representations from these two subspaces via fusing skill to calculation the cross-modal similarity. Compared with previous methods that learning representations on a common space, our model learns mutual information and semantic information in two distinct subspaces, we find our model to be effective to improve the performance and significant gains over GOAT models in of audio-visual cross-modal retrieval task. The result analysis indicates the modality-specific characteristic utilized in the implicit common space, which can be an additional remedy for losing information during bridging the modality gap.
Overall, we emphasize the importance of multi-common spaces during representation learning for audio-visual learning. The comprehensive experiments demonstrate our idea is effective. %The resources for our experiments are available at: https://github.com/dddzeng/　.git.

We plan to propose more powerful fusion skills and try to apply attention mechanisms to consider the local features in implicit space in the future.

\begin{acks}
This work was supported by KDDI research, Inc. Project.
\end{acks}

\bibliographystyle{ACM-Reference-Format}
\bibliography{sample-base}


\begin{thebibliography}{74}

%%% the Web address.

\ifx \showCODEN    \undefined \def \showCODEN     #1{\unskip}     \fi
\ifx \showDOI      \undefined \def \showDOI       #1{#1}\fi
\ifx \showISBNx    \undefined \def \showISBNx     #1{\unskip}     \fi
\ifx \showISBNxiii \undefined \def \showISBNxiii  #1{\unskip}     \fi
\ifx \showISSN     \undefined \def \showISSN      #1{\unskip}     \fi
\ifx \showLCCN     \undefined \def \showLCCN      #1{\unskip}     \fi
\ifx \shownote     \undefined \def \shownote      #1{#1}          \fi
\ifx \showarticletitle \undefined \def \showarticletitle #1{#1}   \fi
\ifx \showURL      \undefined \def \showURL       {\relax}        \fi
% invisible to TeX
\providecommand\bibfield[2]{#2}
\providecommand\bibinfo[2]{#2}
\providecommand\natexlab[1]{#1}
\providecommand\showeprint[2][]{arXiv:#2}

\bibitem[\protect\citeauthoryear{Abu-El-Haija, Kothari, Lee, Natsev, Toderici,
  Varadarajan, and Vijayanarasimhan}{Abu-El-Haija et~al\mbox{.}}{2016}]%
        {abu2016youtube}
\bibfield{author}{\bibinfo{person}{Sami Abu-El-Haija}, \bibinfo{person}{Nisarg
  Kothari}, \bibinfo{person}{Joonseok Lee}, \bibinfo{person}{Paul Natsev},
  \bibinfo{person}{George Toderici}, \bibinfo{person}{Balakrishnan
  Varadarajan}, {and} \bibinfo{person}{Sudheendra Vijayanarasimhan}.}
  \bibinfo{year}{2016}\natexlab{}.
\newblock \bibinfo{title}{YouTube-8M: A Large-Scale Video Classification
  Benchmark}.
\newblock
\newblock
\showeprint[arxiv]{1609.08675}~[cs.CV]


\bibitem[\protect\citeauthoryear{Andrew, Arora, Bilmes, and Livescu}{Andrew
  et~al\mbox{.}}{2013}]%
        {andrew2013deep}
\bibfield{author}{\bibinfo{person}{Galen Andrew}, \bibinfo{person}{Raman
  Arora}, \bibinfo{person}{Jeff Bilmes}, {and} \bibinfo{person}{Karen
  Livescu}.} \bibinfo{year}{2013}\natexlab{}.
\newblock \showarticletitle{Deep Canonical Correlation Analysis}. In
  \bibinfo{booktitle}{\emph{Proceedings of the 30th International Conference on
  Machine Learning}} \emph{(\bibinfo{series}{Proceedings of Machine Learning
  Research}, Vol.~\bibinfo{volume}{28})},
  \bibfield{editor}{\bibinfo{person}{Sanjoy Dasgupta} {and}
  \bibinfo{person}{David McAllester}} (Eds.). \bibinfo{publisher}{PMLR},
  \bibinfo{address}{Atlanta, Georgia, USA}, \bibinfo{pages}{pp.1247--1255}.
\newblock
\urldef\tempurl%
\url{https://proceedings.mlr.press/v28/andrew13.html}
\showURL{%
\tempurl}


\bibitem[\protect\citeauthoryear{Ayyavaraiah and Venkateswarlu}{Ayyavaraiah and
  Venkateswarlu}{2018}]%
        {ayyavaraiah2018joint}
\bibfield{author}{\bibinfo{person}{Monelli Ayyavaraiah} {and}
  \bibinfo{person}{Bondu Venkateswarlu}.} \bibinfo{year}{2018}\natexlab{}.
\newblock \showarticletitle{Joint graph regularization based semantic analysis
  for cross-media retrieval: a systematic review}.
\newblock \bibinfo{journal}{\emph{International Journal of Engineering \&
  Technology}} \bibinfo{volume}{7}, \bibinfo{number}{2.7}
  (\bibinfo{year}{2018}), \bibinfo{pages}{pp.257--261}.
\newblock


\bibitem[\protect\citeauthoryear{Ayyavaraiah and Venkateswarlu}{Ayyavaraiah and
  Venkateswarlu}{2019}]%
        {ayyavaraiah2019cross}
\bibfield{author}{\bibinfo{person}{Monelli Ayyavaraiah} {and}
  \bibinfo{person}{Bondu Venkateswarlu}.} \bibinfo{year}{2019}\natexlab{}.
\newblock \showarticletitle{Cross media feature retrieval and optimization: A
  contemporary review of research scope, challenges and objectives}. In
  \bibinfo{booktitle}{\emph{International Conference On Computational Vision
  and Bio Inspired Computing}}. Springer, \bibinfo{publisher}{Springer
  International Publishing}, \bibinfo{address}{Coimbatore, India},
  \bibinfo{pages}{pp.1125--1136}.
\newblock


\bibitem[\protect\citeauthoryear{Borga}{Borga}{1998}]%
        {borga1998learning}
\bibfield{author}{\bibinfo{person}{Magnus Borga}.}
  \bibinfo{year}{1998}\natexlab{}.
\newblock \emph{\bibinfo{title}{Learning multidimensional signal processing}}.
\newblock \bibinfo{thesistype}{Ph.D. Dissertation}.
  \bibinfo{school}{Link{\"o}ping University Electronic Press}.
\newblock


\bibitem[\protect\citeauthoryear{Bousmalis, Trigeorgis, Silberman, Krishnan,
  and Erhan}{Bousmalis et~al\mbox{.}}{2016}]%
        {bousmalis2016domain}
\bibfield{author}{\bibinfo{person}{Konstantinos Bousmalis},
  \bibinfo{person}{George Trigeorgis}, \bibinfo{person}{Nathan Silberman},
  \bibinfo{person}{Dilip Krishnan}, {and} \bibinfo{person}{Dumitru Erhan}.}
  \bibinfo{year}{2016}\natexlab{}.
\newblock \bibinfo{title}{Domain Separation Networks}.
\newblock
\newblock
\showeprint[arxiv]{1608.06019}~[cs.CV]


\bibitem[\protect\citeauthoryear{Cao, Lin, He, and He}{Cao
  et~al\mbox{.}}{2019}]%
        {CaoLHH19}
\bibfield{author}{\bibinfo{person}{Wenming Cao}, \bibinfo{person}{Qiubin Lin},
  \bibinfo{person}{Zhihai He}, {and} \bibinfo{person}{Zhiquan He}.}
  \bibinfo{year}{2019}\natexlab{}.
\newblock \showarticletitle{Hybrid representation learning for cross-modal
  retrieval}.
\newblock \bibinfo{journal}{\emph{Neurocomputing}}  \bibinfo{volume}{345}
  (\bibinfo{year}{2019}), \bibinfo{pages}{pp.45--57}.
\newblock
\urldef\tempurl%
\url{https://doi.org/10.1016/j.neucom.2018.10.082}
\showDOI{\tempurl}


\bibitem[\protect\citeauthoryear{Cao, Long, Wang, Yang, and Yu}{Cao
  et~al\mbox{.}}{2016}]%
        {CaoLWYY16}
\bibfield{author}{\bibinfo{person}{Yue Cao}, \bibinfo{person}{Mingsheng Long},
  \bibinfo{person}{Jianmin Wang}, \bibinfo{person}{Qiang Yang}, {and}
  \bibinfo{person}{Philip~S. Yu}.} \bibinfo{year}{2016}\natexlab{}.
\newblock \showarticletitle{Deep Visual-Semantic Hashing for Cross-Modal
  Retrieval}. In \bibinfo{booktitle}{\emph{Proceedings of the 22nd {ACM}
  {SIGKDD} International Conference on Knowledge Discovery and Data Mining, San
  Francisco, CA, USA, August 13-17, 2016}},
  \bibfield{editor}{\bibinfo{person}{Balaji Krishnapuram},
  \bibinfo{person}{Mohak Shah}, \bibinfo{person}{Alexander~J. Smola},
  \bibinfo{person}{Charu~C. Aggarwal}, \bibinfo{person}{Dou Shen}, {and}
  \bibinfo{person}{Rajeev Rastogi}} (Eds.). \bibinfo{publisher}{{ACM}},
  \bibinfo{address}{San Francisco, CA, USA}, \bibinfo{pages}{pp.1445--1454}.
\newblock
\urldef\tempurl%
\url{https://doi.org/10.1145/2939672.2939812}
\showDOI{\tempurl}


\bibitem[\protect\citeauthoryear{Carvalho, Cad{\`e}ne, Picard, Soulier, Thome,
  and Cord}{Carvalho et~al\mbox{.}}{2018}]%
        {carvalho2018cross}
\bibfield{author}{\bibinfo{person}{Micael Carvalho}, \bibinfo{person}{R{\'e}mi
  Cad{\`e}ne}, \bibinfo{person}{David Picard}, \bibinfo{person}{Laure Soulier},
  \bibinfo{person}{Nicolas Thome}, {and} \bibinfo{person}{Matthieu Cord}.}
  \bibinfo{year}{2018}\natexlab{}.
\newblock \showarticletitle{Cross-modal retrieval in the cooking context:
  Learning semantic text-image embeddings}. In \bibinfo{booktitle}{\emph{The
  41st International ACM SIGIR Conference on Research \& Development in
  Information Retrieval}}. \bibinfo{publisher}{Association for Computing
  Machinery}, \bibinfo{address}{New York, NY, USA}, \bibinfo{pages}{pp.35--44}.
\newblock


\bibitem[\protect\citeauthoryear{Chaudhuri, Banerjee, Bhattacharya, and
  Datcu}{Chaudhuri et~al\mbox{.}}{2020}]%
        {chaudhuri2020simplified}
\bibfield{author}{\bibinfo{person}{Ushasi Chaudhuri}, \bibinfo{person}{Biplab
  Banerjee}, \bibinfo{person}{Avik Bhattacharya}, {and} \bibinfo{person}{Mihai
  Datcu}.} \bibinfo{year}{2020}\natexlab{}.
\newblock \showarticletitle{A Simplified Framework for Zero-shot Cross-Modal
  Sketch Data Retrieval}. In \bibinfo{booktitle}{\emph{Proceedings of the
  IEEE/CVF Conference on Computer Vision and Pattern Recognition Workshops}}.
  \bibinfo{publisher}{Computer Vision Foundation / {IEEE}},
  \bibinfo{address}{Seattle, WA, USA}, \bibinfo{pages}{pp.182--183}.
\newblock


\bibitem[\protect\citeauthoryear{Chen, Xie, Vedaldi, and Zisserman}{Chen
  et~al\mbox{.}}{2020}]%
        {chen2020vggsound}
\bibfield{author}{\bibinfo{person}{Honglie Chen}, \bibinfo{person}{Weidi Xie},
  \bibinfo{person}{Andrea Vedaldi}, {and} \bibinfo{person}{Andrew Zisserman}.}
  \bibinfo{year}{2020}\natexlab{}.
\newblock \showarticletitle{Vggsound: A large-scale audio-visual dataset}. In
  \bibinfo{booktitle}{\emph{ICASSP 2020-2020 IEEE International Conference on
  Acoustics, Speech and Signal Processing (ICASSP)}}. IEEE,
  \bibinfo{publisher}{IEEE}, \bibinfo{address}{Barcelona, Spain},
  \bibinfo{pages}{pp.721--725}.
\newblock


\bibitem[\protect\citeauthoryear{Chua, Tang, Hong, Li, Luo, and Zheng}{Chua
  et~al\mbox{.}}{2009}]%
        {chua2009nus}
\bibfield{author}{\bibinfo{person}{Tat-Seng Chua}, \bibinfo{person}{Jinhui
  Tang}, \bibinfo{person}{Richang Hong}, \bibinfo{person}{Haojie Li},
  \bibinfo{person}{Zhiping Luo}, {and} \bibinfo{person}{Yantao Zheng}.}
  \bibinfo{year}{2009}\natexlab{}.
\newblock \showarticletitle{Nus-wide: a real-world web image database from
  national university of singapore}. In \bibinfo{booktitle}{\emph{Proceedings
  of the ACM international conference on image and video retrieval}}.
  \bibinfo{publisher}{Association for Computing Machinery},
  \bibinfo{address}{New York, NY, USA}, \bibinfo{pages}{pp.1--9}.
\newblock


\bibitem[\protect\citeauthoryear{Dorfer, Schl{\"u}ter, Vall, Korzeniowski, and
  Widmer}{Dorfer et~al\mbox{.}}{2018}]%
        {dorfer2018end}
\bibfield{author}{\bibinfo{person}{Matthias Dorfer}, \bibinfo{person}{Jan
  Schl{\"u}ter}, \bibinfo{person}{Andreu Vall}, \bibinfo{person}{Filip
  Korzeniowski}, {and} \bibinfo{person}{Gerhard Widmer}.}
  \bibinfo{year}{2018}\natexlab{}.
\newblock \showarticletitle{End-to-end cross-modality retrieval with CCA
  projections and pairwise ranking loss}.
\newblock \bibinfo{journal}{\emph{International Journal of Multimedia
  Information Retrieval}} \bibinfo{volume}{7}, \bibinfo{number}{2}
  (\bibinfo{year}{2018}), \bibinfo{pages}{pp.117--128}.
\newblock


\bibitem[\protect\citeauthoryear{Frome, Corrado, Shlens, Bengio, Dean, Ranzato,
  and Mikolov}{Frome et~al\mbox{.}}{2013}]%
        {FromeCSBDRM13}
\bibfield{author}{\bibinfo{person}{Andrea Frome}, \bibinfo{person}{Gregory~S.
  Corrado}, \bibinfo{person}{Jonathon Shlens}, \bibinfo{person}{Samy Bengio},
  \bibinfo{person}{Jeffrey Dean}, \bibinfo{person}{Marc'Aurelio Ranzato}, {and}
  \bibinfo{person}{Tom{\'{a}}s Mikolov}.} \bibinfo{year}{2013}\natexlab{}.
\newblock \showarticletitle{DeViSE: {A} Deep Visual-Semantic Embedding Model}.
  In \bibinfo{booktitle}{\emph{Advances in Neural Information Processing
  Systems 26: 27th Annual Conference on Neural Information Processing Systems
  2013. Proceedings of a meeting held December 5-8, 2013, Lake Tahoe, Nevada,
  United States}}, \bibfield{editor}{\bibinfo{person}{Christopher J.~C.
  Burges}, \bibinfo{person}{L{\'{e}}on Bottou}, \bibinfo{person}{Zoubin
  Ghahramani}, {and} \bibinfo{person}{Kilian~Q. Weinberger}} (Eds.).
  \bibinfo{publisher}{ACM}, \bibinfo{address}{Lake Tahoe, Nevada, United
  States}, \bibinfo{pages}{pp.2121--2129}.
\newblock
\urldef\tempurl%
\url{https://proceedings.neurips.cc/paper/2013/hash/7cce53cf90577442771720a370c3c723-Abstract.html}
\showURL{%
\tempurl}


\bibitem[\protect\citeauthoryear{Geigle, Pfeiffer, Reimers, Vuli\'{c}, and
  Gurevych}{Geigle et~al\mbox{.}}{2021}]%
        {Geigle2021RetrieveFR}
\bibfield{author}{\bibinfo{person}{Gregor Geigle}, \bibinfo{person}{Jonas
  Pfeiffer}, \bibinfo{person}{Nils Reimers}, \bibinfo{person}{Ivan Vuli\'{c}},
  {and} \bibinfo{person}{Iryna Gurevych}.} \bibinfo{year}{2021}\natexlab{}.
\newblock \showarticletitle{Retrieve Fast, Rerank Smart: Cooperative and Joint
  Approaches for Improved Cross-Modal Retrieval}.
\newblock \bibinfo{journal}{\emph{arXiv preprint}}
  \bibinfo{volume}{abs/2103.11920} (\bibinfo{year}{2021}).
\newblock
\showeprint[arxiv]{2103.11920}
\urldef\tempurl%
\url{http://arxiv.org/abs/2103.11920}
\showURL{%
\tempurl}


\bibitem[\protect\citeauthoryear{Gu, Gu, Gu, Li, Xiong, and Wang}{Gu
  et~al\mbox{.}}{2019}]%
        {GuGGLXW19}
\bibfield{author}{\bibinfo{person}{Wen Gu}, \bibinfo{person}{Xiaoyan Gu},
  \bibinfo{person}{Jingzi Gu}, \bibinfo{person}{Bo Li}, \bibinfo{person}{Zhi
  Xiong}, {and} \bibinfo{person}{Weiping Wang}.}
  \bibinfo{year}{2019}\natexlab{}.
\newblock \showarticletitle{Adversary Guided Asymmetric Hashing for Cross-Modal
  Retrieval}. In \bibinfo{booktitle}{\emph{Proceedings of the 2019 on
  International Conference on Multimedia Retrieval, {ICMR} 2019, Ottawa, ON,
  Canada, June 10-13, 2019.}} \bibinfo{publisher}{Association for Computing
  Machinery}, \bibinfo{address}{New York, NY, USA},
  \bibinfo{pages}{pp.159--167}.
\newblock
\urldef\tempurl%
\url{https://doi.org/10.1145/3323873.3325045}
\showDOI{\tempurl}


\bibitem[\protect\citeauthoryear{Harada, Hayashi, and Uchida}{Harada
  et~al\mbox{.}}{2019}]%
        {harada2019biosignal}
\bibfield{author}{\bibinfo{person}{Shota Harada}, \bibinfo{person}{Hideaki
  Hayashi}, {and} \bibinfo{person}{Seiichi Uchida}.}
  \bibinfo{year}{2019}\natexlab{}.
\newblock \showarticletitle{Biosignal generation and latent variable analysis
  with recurrent generative adversarial networks}.
\newblock \bibinfo{journal}{\emph{IEEE Access}}  \bibinfo{volume}{7}
  (\bibinfo{year}{2019}), \bibinfo{pages}{pp.144292--144302}.
\newblock


\bibitem[\protect\citeauthoryear{Hardoon, Szedm{\'{a}}k, and
  Shawe{-}Taylor}{Hardoon et~al\mbox{.}}{2004}]%
        {hardoon2004canonical}
\bibfield{author}{\bibinfo{person}{David~R. Hardoon},
  \bibinfo{person}{S{\'{a}}ndor Szedm{\'{a}}k}, {and} \bibinfo{person}{John
  Shawe{-}Taylor}.} \bibinfo{year}{2004}\natexlab{}.
\newblock \showarticletitle{Canonical Correlation Analysis: An Overview with
  Application to Learning Methods}.
\newblock \bibinfo{journal}{\emph{Neural Computation.}}
  \bibinfo{volume}{Vol.16}, \bibinfo{number}{no.12} (\bibinfo{year}{2004}),
  \bibinfo{pages}{pp.2639--2664}.
\newblock
\urldef\tempurl%
\url{https://doi.org/10.1162/0899766042321814}
\showDOI{\tempurl}


\bibitem[\protect\citeauthoryear{Hazarika, Zimmermann, and Poria}{Hazarika
  et~al\mbox{.}}{2020}]%
        {hazarika2020misa}
\bibfield{author}{\bibinfo{person}{Devamanyu Hazarika}, \bibinfo{person}{Roger
  Zimmermann}, {and} \bibinfo{person}{Soujanya Poria}.}
  \bibinfo{year}{2020}\natexlab{}.
\newblock \showarticletitle{MISA: Modality-Invariant and-Specific
  Representations for Multimodal Sentiment Analysis}. In
  \bibinfo{booktitle}{\emph{Proceedings of the 28th ACM International
  Conference on Multimedia}}. \bibinfo{publisher}{Association for Computing
  Machinery}, \bibinfo{address}{New York, NY, USA},
  \bibinfo{pages}{pp.1122--1131}.
\newblock


\bibitem[\protect\citeauthoryear{He, Xu, Lu, Yang, Shen, and Shen}{He
  et~al\mbox{.}}{2017}]%
        {XLYSS17}
\bibfield{author}{\bibinfo{person}{Li He}, \bibinfo{person}{Xing Xu},
  \bibinfo{person}{Huimin Lu}, \bibinfo{person}{Yang Yang},
  \bibinfo{person}{Fumin Shen}, {and} \bibinfo{person}{Heng~Tao Shen}.}
  \bibinfo{year}{2017}\natexlab{}.
\newblock \showarticletitle{Unsupervised cross-modal retrieval through
  adversarial learning}. In \bibinfo{booktitle}{\emph{2017 {IEEE} International
  Conference on Multimedia and Expo, {ICME} 2017, Hong Kong, China, July 10-14,
  2017}}. \bibinfo{publisher}{{IEEE} Computer Society}, \bibinfo{address}{Hong
  Kong, China}, \bibinfo{pages}{pp.1153--1158}.
\newblock
\urldef\tempurl%
\url{https://doi.org/10.1109/ICME.2017.8019549}
\showDOI{\tempurl}


\bibitem[\protect\citeauthoryear{Hershey, Chaudhuri, Ellis, Gemmeke, Jansen,
  Moore, Plakal, Platt, Saurous, Seybold, et~al\mbox{.}}{Hershey
  et~al\mbox{.}}{2017}]%
        {hershey2017cnn}
\bibfield{author}{\bibinfo{person}{Shawn Hershey}, \bibinfo{person}{Sourish
  Chaudhuri}, \bibinfo{person}{Daniel~PW Ellis}, \bibinfo{person}{Jort~F
  Gemmeke}, \bibinfo{person}{Aren Jansen}, \bibinfo{person}{R~Channing Moore},
  \bibinfo{person}{Manoj Plakal}, \bibinfo{person}{Devin Platt},
  \bibinfo{person}{Rif~A Saurous}, \bibinfo{person}{Bryan Seybold},
  {et~al\mbox{.}}} \bibinfo{year}{2017}\natexlab{}.
\newblock \showarticletitle{CNN architectures for large-scale audio
  classification}. In \bibinfo{booktitle}{\emph{2017 ieee international
  conference on acoustics, speech and signal processing (icassp)}}. IEEE,
  \bibinfo{publisher}{{IEEE}}, \bibinfo{address}{New Orleans, LA, USA},
  \bibinfo{pages}{pp.131--135}.
\newblock


\bibitem[\protect\citeauthoryear{Hsieh}{Hsieh}{2000}]%
        {hsieh2000nonlinear}
\bibfield{author}{\bibinfo{person}{William~W Hsieh}.}
  \bibinfo{year}{2000}\natexlab{}.
\newblock \showarticletitle{Nonlinear canonical correlation analysis by neural
  networks}.
\newblock \bibinfo{journal}{\emph{Neural Networks}} \bibinfo{volume}{13},
  \bibinfo{number}{10} (\bibinfo{year}{2000}), \bibinfo{pages}{pp.1095--1105}.
\newblock


\bibitem[\protect\citeauthoryear{Jiang, Wu, Li, Zhao, Lu, Tang, and
  Zhuang}{Jiang et~al\mbox{.}}{2015}]%
        {JiangWLZLTZ15}
\bibfield{author}{\bibinfo{person}{Xinyang Jiang}, \bibinfo{person}{Fei Wu},
  \bibinfo{person}{Xi Li}, \bibinfo{person}{Zhou Zhao},
  \bibinfo{person}{Weiming Lu}, \bibinfo{person}{Siliang Tang}, {and}
  \bibinfo{person}{Yueting Zhuang}.} \bibinfo{year}{2015}\natexlab{}.
\newblock \showarticletitle{Deep Compositional Cross-modal Learning to Rank via
  Local-Global Alignment}. In \bibinfo{booktitle}{\emph{Proceedings of the 23rd
  Annual {ACM} Conference on Multimedia Conference, {MM} '15, Brisbane,
  Australia, October 26 - 30, 2015}},
  \bibfield{editor}{\bibinfo{person}{Xiaofang Zhou}, \bibinfo{person}{Alan~F.
  Smeaton}, \bibinfo{person}{Qi~Tian}, \bibinfo{person}{Dick C.~A. Bulterman},
  \bibinfo{person}{Heng~Tao Shen}, \bibinfo{person}{Ketan Mayer{-}Patel}, {and}
  \bibinfo{person}{Shuicheng Yan}} (Eds.). \bibinfo{publisher}{{ACM}},
  \bibinfo{address}{Brisbane, Australia}, \bibinfo{pages}{pp.69--78}.
\newblock
\urldef\tempurl%
\url{https://doi.org/10.1145/2733373.2806240}
\showDOI{\tempurl}


\bibitem[\protect\citeauthoryear{Karpathy, Joulin, and Fei{-}Fei}{Karpathy
  et~al\mbox{.}}{2014}]%
        {KarpathyJL14}
\bibfield{author}{\bibinfo{person}{Andrej Karpathy}, \bibinfo{person}{Armand
  Joulin}, {and} \bibinfo{person}{Li Fei{-}Fei}.}
  \bibinfo{year}{2014}\natexlab{}.
\newblock \showarticletitle{Deep Fragment Embeddings for Bidirectional Image
  Sentence Mapping}. In \bibinfo{booktitle}{\emph{Advances in Neural
  Information Processing Systems 27: Annual Conference on Neural Information
  Processing Systems 2014, December 8-13 2014, Montreal, Quebec, Canada}},
  \bibfield{editor}{\bibinfo{person}{Zoubin Ghahramani}, \bibinfo{person}{Max
  Welling}, \bibinfo{person}{Corinna Cortes}, \bibinfo{person}{Neil~D.
  Lawrence}, {and} \bibinfo{person}{Kilian~Q. Weinberger}} (Eds.).
  \bibinfo{publisher}{ACM}, \bibinfo{address}{Montreal, Quebec, Canada},
  \bibinfo{pages}{pp.1889--1897}.
\newblock
\urldef\tempurl%
\url{https://proceedings.neurips.cc/paper/2014/hash/84d2004bf28a2095230e8e14993d398d-Abstract.html}
\showURL{%
\tempurl}


\bibitem[\protect\citeauthoryear{Kaur, Pannu, and Malhi}{Kaur
  et~al\mbox{.}}{2021}]%
        {kaur2021comparative}
\bibfield{author}{\bibinfo{person}{Parminder Kaur},
  \bibinfo{person}{Husanbir~Singh Pannu}, {and} \bibinfo{person}{Avleen~Kaur
  Malhi}.} \bibinfo{year}{2021}\natexlab{}.
\newblock \showarticletitle{Comparative analysis on cross-modal information
  retrieval: A review}.
\newblock \bibinfo{journal}{\emph{Computer Science Review}}
  \bibinfo{volume}{39} (\bibinfo{year}{2021}), \bibinfo{pages}{pp.100336}.
\newblock


\bibitem[\protect\citeauthoryear{Kay, Carreira, Simonyan, Zhang, Hillier,
  Vijayanarasimhan, Viola, Green, Back, Natsev, Suleyman, and Zisserman}{Kay
  et~al\mbox{.}}{2017}]%
        {will33824}
\bibfield{author}{\bibinfo{person}{Will Kay}, \bibinfo{person}{Jo{\~{a}}o
  Carreira}, \bibinfo{person}{Karen Simonyan}, \bibinfo{person}{Brian Zhang},
  \bibinfo{person}{Chloe Hillier}, \bibinfo{person}{Sudheendra
  Vijayanarasimhan}, \bibinfo{person}{Fabio Viola}, \bibinfo{person}{Tim
  Green}, \bibinfo{person}{Trevor Back}, \bibinfo{person}{Paul Natsev},
  \bibinfo{person}{Mustafa Suleyman}, {and} \bibinfo{person}{Andrew
  Zisserman}.} \bibinfo{year}{2017}\natexlab{}.
\newblock \showarticletitle{The Kinetics Human Action Video Dataset}.
\newblock \bibinfo{journal}{\emph{CoRR}}  \bibinfo{volume}{abs/1705.06950}
  (\bibinfo{year}{2017}).
\newblock
\showeprint[arXiv]{1705.06950}
\urldef\tempurl%
\url{http://arxiv.org/abs/1705.06950}
\showURL{%
\tempurl}


\bibitem[\protect\citeauthoryear{Lai and Fyfe}{Lai and Fyfe}{2000}]%
        {akaho2006kernel}
\bibfield{author}{\bibinfo{person}{Pei~Ling Lai} {and} \bibinfo{person}{Colin
  Fyfe}.} \bibinfo{year}{2000}\natexlab{}.
\newblock \showarticletitle{Kernel and Nonlinear Canonical Correlation
  Analysis}.
\newblock \bibinfo{journal}{\emph{Int. J. Neural Syst.}}
  \bibinfo{volume}{Vol.10}, \bibinfo{number}{no.5} (\bibinfo{year}{2000}),
  \bibinfo{pages}{pp.365--377}.
\newblock
\urldef\tempurl%
\url{https://doi.org/10.1142/S012906570000034X}
\showDOI{\tempurl}


\bibitem[\protect\citeauthoryear{Li, Li, Lu, Zhang, Yin, and Zhang}{Li
  et~al\mbox{.}}{2020}]%
        {LiLLZYZ20}
\bibfield{author}{\bibinfo{person}{Jinxing Li}, \bibinfo{person}{Mu Li},
  \bibinfo{person}{Guangming Lu}, \bibinfo{person}{Bob Zhang},
  \bibinfo{person}{Hongpeng Yin}, {and} \bibinfo{person}{David Zhang}.}
  \bibinfo{year}{2020}\natexlab{}.
\newblock \showarticletitle{Similarity and diversity induced paired projection
  for cross-modal retrieval}.
\newblock \bibinfo{journal}{\emph{Inf. Sci.}}  \bibinfo{volume}{539}
  (\bibinfo{year}{2020}), \bibinfo{pages}{pp.215--228}.
\newblock
\urldef\tempurl%
\url{https://doi.org/10.1016/j.ins.2020.06.032}
\showDOI{\tempurl}


\bibitem[\protect\citeauthoryear{Liu, Qiu, and Huang}{Liu
  et~al\mbox{.}}{2017}]%
        {liu2017adversarial}
\bibfield{author}{\bibinfo{person}{Pengfei Liu}, \bibinfo{person}{Xipeng Qiu},
  {and} \bibinfo{person}{Xuanjing Huang}.} \bibinfo{year}{2017}\natexlab{}.
\newblock \showarticletitle{Adversarial Multi-task Learning for Text
  Classification}. In \bibinfo{booktitle}{\emph{Proceedings of the 55th Annual
  Meeting of the Association for Computational Linguistics, {ACL} 2017,
  Vancouver, Canada, July 30 - August 4, Volume 1: Long Papers}},
  \bibfield{editor}{\bibinfo{person}{Regina Barzilay} {and}
  \bibinfo{person}{Min{-}Yen Kan}} (Eds.). \bibinfo{publisher}{Association for
  Computational Linguistics}, \bibinfo{address}{Vancouver, Canada},
  \bibinfo{pages}{pp.1--10}.
\newblock
\urldef\tempurl%
\url{https://doi.org/10.18653/v1/P17-1001}
\showDOI{\tempurl}


\bibitem[\protect\citeauthoryear{Liu, Nie, Sun, Cui, and Yin}{Liu
  et~al\mbox{.}}{2018}]%
        {liu2018modality}
\bibfield{author}{\bibinfo{person}{Xingbo Liu}, \bibinfo{person}{Xiushan Nie},
  \bibinfo{person}{Haoliang Sun}, \bibinfo{person}{Chaoran Cui}, {and}
  \bibinfo{person}{Yilong Yin}.} \bibinfo{year}{2018}\natexlab{}.
\newblock \showarticletitle{Modality-specific structure preserving hashing for
  cross-modal retrieval}. In \bibinfo{booktitle}{\emph{2018 IEEE International
  Conference on Acoustics, Speech and Signal Processing (ICASSP)}}.
  \bibinfo{publisher}{IEEE}, \bibinfo{address}{Calgary, AB, Canada},
  \bibinfo{pages}{pp.1678--1682}.
\newblock
\urldef\tempurl%
\url{https://doi.org/10.1109/ICASSP.2018.8462454}
\showDOI{\tempurl}


\bibitem[\protect\citeauthoryear{Lu, Duan, and Zhang}{Lu et~al\mbox{.}}{2018}]%
        {lu2018listen}
\bibfield{author}{\bibinfo{person}{Rui Lu}, \bibinfo{person}{Zhiyao Duan},
  {and} \bibinfo{person}{Changshui Zhang}.} \bibinfo{year}{2018}\natexlab{}.
\newblock \showarticletitle{Listen and look: Audio--visual matching assisted
  speech source separation}.
\newblock \bibinfo{journal}{\emph{IEEE Signal Processing Letters}}
  \bibinfo{volume}{25}, \bibinfo{number}{9} (\bibinfo{year}{2018}),
  \bibinfo{pages}{pp.1315--1319}.
\newblock


\bibitem[\protect\citeauthoryear{Ma, Zhang, and Xu}{Ma et~al\mbox{.}}{2020}]%
        {ma2020multi}
\bibfield{author}{\bibinfo{person}{Xinhong Ma}, \bibinfo{person}{Tianzhu
  Zhang}, {and} \bibinfo{person}{Changsheng Xu}.}
  \bibinfo{year}{2020}\natexlab{}.
\newblock \showarticletitle{Multi-level correlation adversarial hashing for
  cross-modal retrieval}.
\newblock \bibinfo{journal}{\emph{IEEE Transactions on Multimedia}}
  \bibinfo{volume}{22}, \bibinfo{number}{12} (\bibinfo{year}{2020}),
  \bibinfo{pages}{pp.3101--3114}.
\newblock


\bibitem[\protect\citeauthoryear{Menon, Surian, and Chawla}{Menon
  et~al\mbox{.}}{2015}]%
        {MenonSC15}
\bibfield{author}{\bibinfo{person}{Aditya~Krishna Menon}, \bibinfo{person}{Didi
  Surian}, {and} \bibinfo{person}{Sanjay Chawla}.}
  \bibinfo{year}{2015}\natexlab{}.
\newblock \showarticletitle{Cross-Modal Retrieval: {A} Pairwise Classification
  Approach}. In \bibinfo{booktitle}{\emph{Proceedings of the 2015 {SIAM}
  International Conference on Data Mining, Vancouver, BC, Canada, April 30 -
  May 2, 2015}}, \bibfield{editor}{\bibinfo{person}{Suresh Venkatasubramanian}
  {and} \bibinfo{person}{Jieping Ye}} (Eds.). \bibinfo{publisher}{{SIAM}},
  \bibinfo{address}{Vancouver, BC, Canada}, \bibinfo{pages}{pp.199--207}.
\newblock
\urldef\tempurl%
\url{https://doi.org/10.1137/1.9781611974010.23}
\showDOI{\tempurl}


\bibitem[\protect\citeauthoryear{M{\"u}ller, Arzt, Balke, Dorfer, and
  Widmer}{M{\"u}ller et~al\mbox{.}}{2018}]%
        {muller2018cross}
\bibfield{author}{\bibinfo{person}{Meinard M{\"u}ller},
  \bibinfo{person}{Andreas Arzt}, \bibinfo{person}{Stefan Balke},
  \bibinfo{person}{Matthias Dorfer}, {and} \bibinfo{person}{Gerhard Widmer}.}
  \bibinfo{year}{2018}\natexlab{}.
\newblock \showarticletitle{Cross-modal music retrieval and applications: An
  overview of key methodologies}.
\newblock \bibinfo{journal}{\emph{IEEE Signal Processing Magazine}}
  \bibinfo{volume}{36}, \bibinfo{number}{1} (\bibinfo{year}{2018}),
  \bibinfo{pages}{pp.52--62}.
\newblock


\bibitem[\protect\citeauthoryear{Ngiam, Khosla, Kim, Nam, Lee, and Ng}{Ngiam
  et~al\mbox{.}}{2011}]%
        {ngiam2011multimodal}
\bibfield{author}{\bibinfo{person}{Jiquan Ngiam}, \bibinfo{person}{Aditya
  Khosla}, \bibinfo{person}{Mingyu Kim}, \bibinfo{person}{Juhan Nam},
  \bibinfo{person}{Honglak Lee}, {and} \bibinfo{person}{Andrew~Y. Ng}.}
  \bibinfo{year}{2011}\natexlab{}.
\newblock \showarticletitle{Multimodal Deep Learning}. In
  \bibinfo{booktitle}{\emph{Proceedings of the 28th International Conference on
  Machine Learning, {ICML} 2011, Bellevue, Washington, USA, June 28 - July 2,
  2011}}. \bibinfo{publisher}{Omnipress}, \bibinfo{address}{Madison, WI, USA},
  \bibinfo{pages}{pp.689--696}.
\newblock


\bibitem[\protect\citeauthoryear{Nie, Wang, Li, Hao, Jian, and Yin}{Nie
  et~al\mbox{.}}{2021}]%
        {NieWLHJY21}
\bibfield{author}{\bibinfo{person}{Xiushan Nie}, \bibinfo{person}{Bowei Wang},
  \bibinfo{person}{Jiajia Li}, \bibinfo{person}{Fanchang Hao},
  \bibinfo{person}{Muwei Jian}, {and} \bibinfo{person}{Yilong Yin}.}
  \bibinfo{year}{2021}\natexlab{}.
\newblock \showarticletitle{Deep Multiscale Fusion Hashing for Cross-Modal
  Retrieval}.
\newblock \bibinfo{journal}{\emph{{IEEE} Trans. Circuits Syst. Video Technol.}}
  \bibinfo{volume}{31}, \bibinfo{number}{1} (\bibinfo{year}{2021}),
  \bibinfo{pages}{pp.401--410}.
\newblock
\urldef\tempurl%
\url{https://doi.org/10.1109/TCSVT.2020.2974877}
\showDOI{\tempurl}


\bibitem[\protect\citeauthoryear{Peng, Huang, and Zhao}{Peng
  et~al\mbox{.}}{2017a}]%
        {peng2017overview}
\bibfield{author}{\bibinfo{person}{Yuxin Peng}, \bibinfo{person}{Xin Huang},
  {and} \bibinfo{person}{Yunzhen Zhao}.} \bibinfo{year}{2017}\natexlab{a}.
\newblock \showarticletitle{An overview of cross-media retrieval: Concepts,
  methodologies, benchmarks, and challenges}.
\newblock \bibinfo{journal}{\emph{IEEE Transactions on circuits and systems for
  video technology}} \bibinfo{volume}{28}, \bibinfo{number}{9}
  (\bibinfo{year}{2017}), \bibinfo{pages}{pp.2372--2385}.
\newblock


\bibitem[\protect\citeauthoryear{Peng, Qi, and Yuan}{Peng
  et~al\mbox{.}}{2018a}]%
        {peng2018modality}
\bibfield{author}{\bibinfo{person}{Yuxin Peng}, \bibinfo{person}{Jinwei Qi},
  {and} \bibinfo{person}{Yuxin Yuan}.} \bibinfo{year}{2018}\natexlab{a}.
\newblock \showarticletitle{Modality-specific cross-modal similarity
  measurement with recurrent attention network}.
\newblock \bibinfo{journal}{\emph{IEEE Transactions on Image Processing}}
  \bibinfo{volume}{27}, \bibinfo{number}{11} (\bibinfo{year}{2018}),
  \bibinfo{pages}{pp.5585--5599}.
\newblock


\bibitem[\protect\citeauthoryear{Peng, Qi, and Yuan}{Peng
  et~al\mbox{.}}{2018b}]%
        {PengQY18}
\bibfield{author}{\bibinfo{person}{Yuxin Peng}, \bibinfo{person}{Jinwei Qi},
  {and} \bibinfo{person}{Yuxin Yuan}.} \bibinfo{year}{2018}\natexlab{b}.
\newblock \showarticletitle{Modality-Specific Cross-Modal Similarity
  Measurement With Recurrent Attention Network}.
\newblock \bibinfo{journal}{\emph{{IEEE} Trans. Image Process.}}
  \bibinfo{volume}{27}, \bibinfo{number}{11} (\bibinfo{year}{2018}),
  \bibinfo{pages}{pp.5585--5599}.
\newblock
\urldef\tempurl%
\url{https://doi.org/10.1109/TIP.2018.2852503}
\showDOI{\tempurl}


\bibitem[\protect\citeauthoryear{Peng, Zhu, Zhao, Xu, Huang, Lu, Zheng, Huang,
  and Gao}{Peng et~al\mbox{.}}{2017b}]%
        {peng2017cross}
\bibfield{author}{\bibinfo{person}{Yu-xin Peng}, \bibinfo{person}{Wen-wu Zhu},
  \bibinfo{person}{Yao Zhao}, \bibinfo{person}{Chang-sheng Xu},
  \bibinfo{person}{Qing-ming Huang}, \bibinfo{person}{Han-qing Lu},
  \bibinfo{person}{Qing-hua Zheng}, \bibinfo{person}{Tie-jun Huang}, {and}
  \bibinfo{person}{Wen Gao}.} \bibinfo{year}{2017}\natexlab{b}.
\newblock \showarticletitle{Cross-media analysis and reasoning: advances and
  directions}.
\newblock \bibinfo{journal}{\emph{Frontiers of Information Technology \&
  Electronic Engineering}} \bibinfo{volume}{18}, \bibinfo{number}{1}
  (\bibinfo{year}{2017}), \bibinfo{pages}{pp.44--57}.
\newblock


\bibitem[\protect\citeauthoryear{Pereira, Coviello, Doyle, Rasiwasia,
  Lanckriet, Levy, and Vasconcelos}{Pereira et~al\mbox{.}}{2013}]%
        {pereira2013role}
\bibfield{author}{\bibinfo{person}{Jose~Costa Pereira},
  \bibinfo{person}{Emanuele Coviello}, \bibinfo{person}{Gabriel Doyle},
  \bibinfo{person}{Nikhil Rasiwasia}, \bibinfo{person}{Gert~RG Lanckriet},
  \bibinfo{person}{Roger Levy}, {and} \bibinfo{person}{Nuno Vasconcelos}.}
  \bibinfo{year}{2013}\natexlab{}.
\newblock \showarticletitle{On the role of correlation and abstraction in
  cross-modal multimedia retrieval}.
\newblock \bibinfo{journal}{\emph{IEEE transactions on pattern analysis and
  machine intelligence}} \bibinfo{volume}{36}, \bibinfo{number}{3}
  (\bibinfo{year}{2013}), \bibinfo{pages}{pp.521--535}.
\newblock


\bibitem[\protect\citeauthoryear{Priyanka, Devi, Riyazoddin, and
  Reddy}{Priyanka et~al\mbox{.}}{2013}]%
        {priyanka2013analysis}
\bibfield{author}{\bibinfo{person}{M Priyanka}, \bibinfo{person}{B~Sunita
  Devi}, \bibinfo{person}{SM Riyazoddin}, {and} \bibinfo{person}{M~Janga
  Reddy}.} \bibinfo{year}{2013}\natexlab{}.
\newblock \showarticletitle{Analysis of cross-media web information fusion for
  text and image association-a survey paper}.
\newblock \bibinfo{journal}{\emph{Global Journal of Computer Science and
  Technology}} \bibinfo{volume}{13}, \bibinfo{number}{1}
  (\bibinfo{year}{2013}), \bibinfo{pages}{pp.9--15}.
\newblock


\bibitem[\protect\citeauthoryear{Ranjan, Rasiwasia, and Jawahar}{Ranjan
  et~al\mbox{.}}{2015}]%
        {ranjan2015multi}
\bibfield{author}{\bibinfo{person}{Viresh Ranjan}, \bibinfo{person}{Nikhil
  Rasiwasia}, {and} \bibinfo{person}{CV Jawahar}.}
  \bibinfo{year}{2015}\natexlab{}.
\newblock \showarticletitle{Multi-label cross-modal retrieval}. In
  \bibinfo{booktitle}{\emph{Proceedings of the IEEE International Conference on
  Computer Vision}}. \bibinfo{publisher}{{IEEE} Computer Society},
  \bibinfo{address}{Santiago, Chile}, \bibinfo{pages}{pp.4094--4102}.
\newblock


\bibitem[\protect\citeauthoryear{Rasiwasia, Mahajan, Mahadevan, and
  Aggarwal}{Rasiwasia et~al\mbox{.}}{2014}]%
        {rasiwasia2014cluster}
\bibfield{author}{\bibinfo{person}{Nikhil Rasiwasia}, \bibinfo{person}{Dhruv
  Mahajan}, \bibinfo{person}{Vijay Mahadevan}, {and} \bibinfo{person}{Gaurav
  Aggarwal}.} \bibinfo{year}{2014}\natexlab{}.
\newblock \showarticletitle{Cluster Canonical Correlation Analysis}. In
  \bibinfo{booktitle}{\emph{Proceedings of the Seventeenth International
  Conference on Artificial Intelligence and Statistics, {AISTATS} 2014,
  Reykjavik, Iceland, April 22-25, 2014}}. \bibinfo{publisher}{JMLR.org},
  \bibinfo{address}{Reykjavik, Iceland}, \bibinfo{pages}{pp.823--831}.
\newblock
\urldef\tempurl%
\url{https://doi.org/10.1201/b18358-8}
\showDOI{\tempurl}


\bibitem[\protect\citeauthoryear{Roy, Verma, Ghosh, and Ghosh}{Roy
  et~al\mbox{.}}{2020}]%
        {roy2020zscrgan}
\bibfield{author}{\bibinfo{person}{Anurag Roy}, \bibinfo{person}{Vinay~Kumar
  Verma}, \bibinfo{person}{Kripabandhu Ghosh}, {and} \bibinfo{person}{Saptarshi
  Ghosh}.} \bibinfo{year}{2020}\natexlab{}.
\newblock \showarticletitle{ZSCRGAN: A GAN-based Expectation Maximization Model
  for Zero-Shot Retrieval of Images from Textual Descriptions}. In
  \bibinfo{booktitle}{\emph{Proceedings of the 29th ACM International
  Conference on Information \& Knowledge Management}}.
  \bibinfo{publisher}{{ACM}}, \bibinfo{address}{Virtual Event, Ireland},
  \bibinfo{pages}{pp.1315--1324}.
\newblock


\bibitem[\protect\citeauthoryear{Ruder and Plank}{Ruder and Plank}{2018}]%
        {ruder2018strong}
\bibfield{author}{\bibinfo{person}{Sebastian Ruder} {and}
  \bibinfo{person}{Barbara Plank}.} \bibinfo{year}{2018}\natexlab{}.
\newblock \showarticletitle{Strong Baselines for Neural Semi-Supervised
  Learning under Domain Shift}. In \bibinfo{booktitle}{\emph{Proceedings of the
  56th Annual Meeting of the Association for Computational Linguistics, {ACL}
  2018, Melbourne, Australia, July 15-20, 2018, Volume 1: Long Papers}},
  \bibfield{editor}{\bibinfo{person}{Iryna Gurevych} {and}
  \bibinfo{person}{Yusuke Miyao}} (Eds.). \bibinfo{publisher}{Association for
  Computational Linguistics}, \bibinfo{address}{Melbourne, Australia},
  \bibinfo{pages}{pp.1044--1054}.
\newblock
\urldef\tempurl%
\url{https://doi.org/10.18653/v1/P18-1096}
\showDOI{\tempurl}


\bibitem[\protect\citeauthoryear{Shang, Zhang, Zhu, and Sun}{Shang
  et~al\mbox{.}}{2019}]%
        {ShangZZS19}
\bibfield{author}{\bibinfo{person}{Fei Shang}, \bibinfo{person}{Huaxiang
  Zhang}, \bibinfo{person}{Lei Zhu}, {and} \bibinfo{person}{Jiande Sun}.}
  \bibinfo{year}{2019}\natexlab{}.
\newblock \showarticletitle{Adversarial cross-modal retrieval based on
  dictionary learning}.
\newblock \bibinfo{journal}{\emph{Neurocomputing}}  \bibinfo{volume}{355}
  (\bibinfo{year}{2019}), \bibinfo{pages}{pp.93--104}.
\newblock
\urldef\tempurl%
\url{https://doi.org/10.1016/j.neucom.2019.04.041}
\showDOI{\tempurl}


\bibitem[\protect\citeauthoryear{Shao, Zhao, Su, and Yue}{Shao
  et~al\mbox{.}}{2017}]%
        {ShaoZSY17}
\bibfield{author}{\bibinfo{person}{Jie Shao}, \bibinfo{person}{Zhicheng Zhao},
  \bibinfo{person}{Fei Su}, {and} \bibinfo{person}{Ting Yue}.}
  \bibinfo{year}{2017}\natexlab{}.
\newblock \showarticletitle{Towards Improving Canonical Correlation Analysis
  for Cross-modal Retrieval}. In \bibinfo{booktitle}{\emph{Proceedings of the
  on Thematic Workshops of {ACM} Multimedia 2017, Mountain View, CA, USA,
  October 23 - 27, 2017}}, \bibfield{editor}{\bibinfo{person}{Wanmin Wu},
  \bibinfo{person}{Jianchao Yang}, \bibinfo{person}{Qi~Tian}, {and}
  \bibinfo{person}{Roger Zimmermann}} (Eds.). \bibinfo{publisher}{{ACM}},
  \bibinfo{address}{Mountain View, CA, USA}, \bibinfo{pages}{pp.332--339}.
\newblock
\urldef\tempurl%
\url{https://doi.org/10.1145/3126686.3126726}
\showDOI{\tempurl}


\bibitem[\protect\citeauthoryear{Socher, Karpathy, Le, Manning, and Ng}{Socher
  et~al\mbox{.}}{2014}]%
        {socher2014grounded}
\bibfield{author}{\bibinfo{person}{Richard Socher}, \bibinfo{person}{Andrej
  Karpathy}, \bibinfo{person}{Quoc~V Le}, \bibinfo{person}{Christopher~D
  Manning}, {and} \bibinfo{person}{Andrew~Y Ng}.}
  \bibinfo{year}{2014}\natexlab{}.
\newblock \showarticletitle{Grounded compositional semantics for finding and
  describing images with sentences}.
\newblock \bibinfo{journal}{\emph{Transactions of the Association for
  Computational Linguistics}}  \bibinfo{volume}{2} (\bibinfo{year}{2014}),
  \bibinfo{pages}{pp.207--218}.
\newblock


\bibitem[\protect\citeauthoryear{Tian, Shi, Li, Duan, and Xu}{Tian
  et~al\mbox{.}}{2018}]%
        {tian2018audio}
\bibfield{author}{\bibinfo{person}{Yapeng Tian}, \bibinfo{person}{Jing Shi},
  \bibinfo{person}{Bochen Li}, \bibinfo{person}{Zhiyao Duan}, {and}
  \bibinfo{person}{Chenliang Xu}.} \bibinfo{year}{2018}\natexlab{}.
\newblock \showarticletitle{Audio-Visual Event Localization in Unconstrained
  Videos}. In \bibinfo{booktitle}{\emph{Computer Vision - {ECCV} 2018 - 15th
  European Conference, Munich, Germany, September 8-14, 2018, Proceedings, Part
  {II}}} \emph{(\bibinfo{series}{Lecture Notes in Computer Science},
  Vol.~\bibinfo{volume}{11206})}, \bibfield{editor}{\bibinfo{person}{Vittorio
  Ferrari}, \bibinfo{person}{Martial Hebert}, \bibinfo{person}{Cristian
  Sminchisescu}, {and} \bibinfo{person}{Yair Weiss}} (Eds.).
  \bibinfo{publisher}{Springer}, \bibinfo{address}{Munich, Germany},
  \bibinfo{pages}{pp.252--268}.
\newblock
\urldef\tempurl%
\url{https://doi.org/10.1007/978-3-030-01216-8\_16}
\showDOI{\tempurl}


\bibitem[\protect\citeauthoryear{Wang, Yang, Xu, Hanjalic, and Shen}{Wang
  et~al\mbox{.}}{2017}]%
        {wang2017adversarial}
\bibfield{author}{\bibinfo{person}{Bokun Wang}, \bibinfo{person}{Yang Yang},
  \bibinfo{person}{Xing Xu}, \bibinfo{person}{Alan Hanjalic}, {and}
  \bibinfo{person}{Heng~Tao Shen}.} \bibinfo{year}{2017}\natexlab{}.
\newblock \showarticletitle{Adversarial Cross-Modal Retrieval}. In
  \bibinfo{booktitle}{\emph{Proceedings of the 25th ACM International
  Conference on Multimedia}} (Mountain View, California, USA)
  \emph{(\bibinfo{series}{MM '17})}. \bibinfo{publisher}{Association for
  Computing Machinery}, \bibinfo{address}{New York, NY, USA},
  \bibinfo{pages}{pp.154–162}.
\newblock
\showISBNx{9781450349062}
\urldef\tempurl%
\url{https://doi.org/10.1145/3123266.3123326}
\showDOI{\tempurl}


\bibitem[\protect\citeauthoryear{Wang, He, Kang, Xiang, and Pan}{Wang
  et~al\mbox{.}}{2015}]%
        {WangHKXP15}
\bibfield{author}{\bibinfo{person}{Jian Wang}, \bibinfo{person}{Yonghao He},
  \bibinfo{person}{Cuicui Kang}, \bibinfo{person}{Shiming Xiang}, {and}
  \bibinfo{person}{Chunhong Pan}.} \bibinfo{year}{2015}\natexlab{}.
\newblock \showarticletitle{Image-Text Cross-Modal Retrieval via
  Modality-Specific Feature Learning}. In \bibinfo{booktitle}{\emph{Proceedings
  of the 5th {ACM} on International Conference on Multimedia Retrieval,
  Shanghai, China, June 23-26, 2015}},
  \bibfield{editor}{\bibinfo{person}{Alexander~G. Hauptmann},
  \bibinfo{person}{Chong{-}Wah Ngo}, \bibinfo{person}{Xiangyang Xue},
  \bibinfo{person}{Yu{-}Gang Jiang}, \bibinfo{person}{Cees Snoek}, {and}
  \bibinfo{person}{Nuno Vasconcelos}} (Eds.). \bibinfo{publisher}{{ACM}},
  \bibinfo{address}{Shanghai, China}, \bibinfo{pages}{pp.347--354}.
\newblock
\urldef\tempurl%
\url{https://doi.org/10.1145/2671188.2749341}
\showDOI{\tempurl}


\bibitem[\protect\citeauthoryear{Wang, Yin, Wang, Wu, and Wang}{Wang
  et~al\mbox{.}}{2016b}]%
        {wang2016comprehensive}
\bibfield{author}{\bibinfo{person}{Kaiye Wang}, \bibinfo{person}{Qiyue Yin},
  \bibinfo{person}{Wei Wang}, \bibinfo{person}{Shu Wu}, {and}
  \bibinfo{person}{Liang Wang}.} \bibinfo{year}{2016}\natexlab{b}.
\newblock \showarticletitle{A Comprehensive Survey on Cross-modal Retrieval}.
\newblock \bibinfo{journal}{\emph{CoRR}}  \bibinfo{volume}{abs/1607.06215}
  (\bibinfo{year}{2016}), \bibinfo{pages}{arXiv:1607.06215}.
\newblock
\showeprint[arXiv]{1607.06215}
\urldef\tempurl%
\url{http://arxiv.org/abs/1607.06215}
\showURL{%
\tempurl}


\bibitem[\protect\citeauthoryear{Wang, Yang, Ooi, Zhang, and Zhuang}{Wang
  et~al\mbox{.}}{2016a}]%
        {wang2016effective}
\bibfield{author}{\bibinfo{person}{Wei Wang}, \bibinfo{person}{Xiaoyan Yang},
  \bibinfo{person}{Beng~Chin Ooi}, \bibinfo{person}{Dongxiang Zhang}, {and}
  \bibinfo{person}{Yueting Zhuang}.} \bibinfo{year}{2016}\natexlab{a}.
\newblock \showarticletitle{Effective deep learning-based multi-modal
  retrieval}.
\newblock \bibinfo{journal}{\emph{The VLDB Journal}} \bibinfo{volume}{25},
  \bibinfo{number}{1} (\bibinfo{year}{2016}), \bibinfo{pages}{pp.79--101}.
\newblock


\bibitem[\protect\citeauthoryear{Wu, Jing, Wu, Ji, Dong, Luo, Huang, and
  Wang}{Wu et~al\mbox{.}}{2020}]%
        {wu2020modality}
\bibfield{author}{\bibinfo{person}{Fei Wu}, \bibinfo{person}{Xiao-Yuan Jing},
  \bibinfo{person}{Zhiyong Wu}, \bibinfo{person}{Yimu Ji},
  \bibinfo{person}{Xiwei Dong}, \bibinfo{person}{Xiaokai Luo},
  \bibinfo{person}{Qinghua Huang}, {and} \bibinfo{person}{Ruchuan Wang}.}
  \bibinfo{year}{2020}\natexlab{}.
\newblock \showarticletitle{Modality-specific and shared generative adversarial
  network for cross-modal retrieval}.
\newblock \bibinfo{journal}{\emph{Pattern Recognition}}  \bibinfo{volume}{104}
  (\bibinfo{year}{2020}), \bibinfo{pages}{pp.107335}.
\newblock


\bibitem[\protect\citeauthoryear{Xie, Deng, Li, Liu, and Tao}{Xie
  et~al\mbox{.}}{2020}]%
        {xie2020multi}
\bibfield{author}{\bibinfo{person}{De Xie}, \bibinfo{person}{Cheng Deng},
  \bibinfo{person}{Chao Li}, \bibinfo{person}{Xianglong Liu}, {and}
  \bibinfo{person}{Dacheng Tao}.} \bibinfo{year}{2020}\natexlab{}.
\newblock \showarticletitle{Multi-task consistency-preserving adversarial
  hashing for cross-modal retrieval}.
\newblock \bibinfo{journal}{\emph{IEEE Transactions on Image Processing}}
  \bibinfo{volume}{29} (\bibinfo{year}{2020}), \bibinfo{pages}{pp.3626--3637}.
\newblock


\bibitem[\protect\citeauthoryear{Xiong, Ou, Yan, Gou, Zhou, and Wang}{Xiong
  et~al\mbox{.}}{2020}]%
        {xiong2020modality}
\bibfield{author}{\bibinfo{person}{Haixia Xiong}, \bibinfo{person}{Weihua Ou},
  \bibinfo{person}{Zengxian Yan}, \bibinfo{person}{Jianping Gou},
  \bibinfo{person}{Quan Zhou}, {and} \bibinfo{person}{Anzhi Wang}.}
  \bibinfo{year}{2020}\natexlab{}.
\newblock \showarticletitle{Modality-specific matrix factorization hashing for
  cross-modal retrieval}.
\newblock \bibinfo{journal}{\emph{Journal of Ambient Intelligence and Humanized
  Computing}}  \bibinfo{volume}{1} (\bibinfo{year}{2020}),
  \bibinfo{pages}{pp.1--15}.
\newblock
\urldef\tempurl%
\url{https://doi.org/10.1007/s12652-020-02177-7}
\showDOI{\tempurl}


\bibitem[\protect\citeauthoryear{Xu, Li, and Zhang}{Xu et~al\mbox{.}}{2020}]%
        {XuLZ20}
\bibfield{author}{\bibinfo{person}{Gongwen Xu}, \bibinfo{person}{Xiaomei Li},
  {and} \bibinfo{person}{Zhijun Zhang}.} \bibinfo{year}{2020}\natexlab{}.
\newblock \showarticletitle{Semantic Consistency Cross-Modal Retrieval With
  Semi-Supervised Graph Regularization}.
\newblock \bibinfo{journal}{\emph{{IEEE} Access}}  \bibinfo{volume}{8}
  (\bibinfo{year}{2020}), \bibinfo{pages}{pp.14278--14288}.
\newblock
\urldef\tempurl%
\url{https://doi.org/10.1109/ACCESS.2020.2966220}
\showDOI{\tempurl}


\bibitem[\protect\citeauthoryear{Xu, Lin, Yang, Hanjalic, and Shen}{Xu
  et~al\mbox{.}}{5555}]%
        {xu9296975}
\bibfield{author}{\bibinfo{person}{X. Xu}, \bibinfo{person}{K. Lin},
  \bibinfo{person}{Y. Yang}, \bibinfo{person}{A. Hanjalic}, {and}
  \bibinfo{person}{H. Shen}.} \bibinfo{year}{5555}\natexlab{}.
\newblock \showarticletitle{Joint Feature Synthesis and Embedding: Adversarial
  Cross-modal Retrieval Revisited}.
\newblock \bibinfo{journal}{\emph{IEEE Transactions on Pattern Analysis and
  Machine Intelligence}} \bibinfo{volume}{1}, \bibinfo{number}{01}
  (\bibinfo{date}{dec} \bibinfo{year}{5555}), \bibinfo{pages}{pp.1--1}.
\newblock
\showISSN{1939-3539}
\urldef\tempurl%
\url{https://doi.org/10.1109/TPAMI.2020.3045530}
\showDOI{\tempurl}


\bibitem[\protect\citeauthoryear{{Xu}, {Lu}, {Song}, {Yang}, {Shen}, and
  {Li}}{{Xu} et~al\mbox{.}}{2020}]%
        {ter8771379}
\bibfield{author}{\bibinfo{person}{X. {Xu}}, \bibinfo{person}{H. {Lu}},
  \bibinfo{person}{J. {Song}}, \bibinfo{person}{Y. {Yang}},
  \bibinfo{person}{H.~T. {Shen}}, {and} \bibinfo{person}{X. {Li}}.}
  \bibinfo{year}{2020}\natexlab{}.
\newblock \showarticletitle{Ternary Adversarial Networks With Self-Supervision
  for Zero-Shot Cross-Modal Retrieval}.
\newblock \bibinfo{journal}{\emph{IEEE Transactions on Cybernetics}}
  \bibinfo{volume}{50}, \bibinfo{number}{6} (\bibinfo{year}{2020}),
  \bibinfo{pages}{pp.2400--2413}.
\newblock
\urldef\tempurl%
\url{https://doi.org/10.1109/TCYB.2019.2928180}
\showDOI{\tempurl}


\bibitem[\protect\citeauthoryear{Yanagi, Togo, Ogawa, and Haseyama}{Yanagi
  et~al\mbox{.}}{2020}]%
        {yanagi2020enhancing}
\bibfield{author}{\bibinfo{person}{Rintaro Yanagi}, \bibinfo{person}{Ren Togo},
  \bibinfo{person}{Takahiro Ogawa}, {and} \bibinfo{person}{Miki Haseyama}.}
  \bibinfo{year}{2020}\natexlab{}.
\newblock \showarticletitle{Enhancing cross-modal retrieval based on
  modality-specific and embedding spaces}.
\newblock \bibinfo{journal}{\emph{IEEE Access}}  \bibinfo{volume}{8}
  (\bibinfo{year}{2020}), \bibinfo{pages}{pp.96777--96786}.
\newblock


\bibitem[\protect\citeauthoryear{Yang, Deng, Liu, Liu, Tao, and Gao}{Yang
  et~al\mbox{.}}{2017}]%
        {yang2017pairwise}
\bibfield{author}{\bibinfo{person}{Erkun Yang}, \bibinfo{person}{Cheng Deng},
  \bibinfo{person}{Wei Liu}, \bibinfo{person}{Xianglong Liu},
  \bibinfo{person}{Dacheng Tao}, {and} \bibinfo{person}{Xinbo Gao}.}
  \bibinfo{year}{2017}\natexlab{}.
\newblock \showarticletitle{Pairwise relationship guided deep hashing for
  cross-modal retrieval}. In \bibinfo{booktitle}{\emph{proceedings of the AAAI
  Conference on Artificial Intelligence}}, Vol.~\bibinfo{volume}{31}.
  \bibinfo{publisher}{{AAAI} Press}, \bibinfo{address}{San Francisco,
  California, {USA}}, \bibinfo{pages}{pp.1618--1625}.
\newblock
\urldef\tempurl%
\url{http://aaai.org/ocs/index.php/AAAI/AAAI17/paper/view/14326}
\showURL{%
\tempurl}


\bibitem[\protect\citeauthoryear{Yu, Tang, Aizawa, and Aizawa}{Yu
  et~al\mbox{.}}{2018}]%
        {yu2018category}
\bibfield{author}{\bibinfo{person}{Yi Yu}, \bibinfo{person}{Suhua Tang},
  \bibinfo{person}{Kiyoharu Aizawa}, {and} \bibinfo{person}{Akiko Aizawa}.}
  \bibinfo{year}{2018}\natexlab{}.
\newblock \showarticletitle{Category-based deep CCA for fine-grained venue
  discovery from multimodal data}.
\newblock \bibinfo{journal}{\emph{IEEE transactions on neural networks and
  learning systems}} \bibinfo{volume}{30}, \bibinfo{number}{4}
  (\bibinfo{year}{2018}), \bibinfo{pages}{pp.1250--1258}.
\newblock


\bibitem[\protect\citeauthoryear{Zeng and Oyama}{Zeng and Oyama}{2019}]%
        {zeng2019learning}
\bibfield{author}{\bibinfo{person}{Donghuo Zeng} {and} \bibinfo{person}{Keizo
  Oyama}.} \bibinfo{year}{2019}\natexlab{}.
\newblock \showarticletitle{Learning Joint Embedding for Cross-Modal
  Retrieval}. In \bibinfo{booktitle}{\emph{2019 International Conference on
  Data Mining Workshops (ICDMW)}}. IEEE, \bibinfo{publisher}{IEEE},
  \bibinfo{address}{Bejing, China.}, \bibinfo{pages}{pp.1070--1071}.
\newblock


\bibitem[\protect\citeauthoryear{Zeng, Yu, and Oyama}{Zeng
  et~al\mbox{.}}{2018}]%
        {zeng2018audio}
\bibfield{author}{\bibinfo{person}{Donghuo Zeng}, \bibinfo{person}{Yi Yu},
  {and} \bibinfo{person}{Keizo Oyama}.} \bibinfo{year}{2018}\natexlab{}.
\newblock \showarticletitle{Audio-Visual Embedding for Cross-Modal Music Video
  Retrieval through Supervised Deep {CCA}}. In \bibinfo{booktitle}{\emph{2018
  {IEEE} International Symposium on Multimedia, {ISM} 2018, Taichung, Taiwan,
  December 10-12, 2018}}. \bibinfo{publisher}{{IEEE} Computer Society},
  \bibinfo{address}{Taichung, Taiwan}, \bibinfo{pages}{pp.143--150}.
\newblock
\urldef\tempurl%
\url{https://doi.org/10.1109/ISM.2018.00-21}
\showDOI{\tempurl}


\bibitem[\protect\citeauthoryear{Zeng, Yu, and Oyama}{Zeng
  et~al\mbox{.}}{2020}]%
        {zeng2020deep}
\bibfield{author}{\bibinfo{person}{Donghuo Zeng}, \bibinfo{person}{Yi Yu},
  {and} \bibinfo{person}{Keizo Oyama}.} \bibinfo{year}{2020}\natexlab{}.
\newblock \showarticletitle{Deep triplet neural networks with cluster-cca for
  audio-visual cross-modal retrieval}.
\newblock \bibinfo{journal}{\emph{ACM Transactions on Multimedia Computing,
  Communications, and Applications (TOMM)}} \bibinfo{volume}{16},
  \bibinfo{number}{3} (\bibinfo{year}{2020}), \bibinfo{pages}{pp.1--23}.
\newblock


\bibitem[\protect\citeauthoryear{Zhang, Wu, and Yu}{Zhang
  et~al\mbox{.}}{2021}]%
        {zhang2021label}
\bibfield{author}{\bibinfo{person}{Donglin Zhang}, \bibinfo{person}{Xiao-Jun
  Wu}, {and} \bibinfo{person}{Jun Yu}.} \bibinfo{year}{2021}\natexlab{}.
\newblock \showarticletitle{Label consistent flexible matrix factorization
  hashing for efficient cross-modal retrieval}.
\newblock \bibinfo{journal}{\emph{ACM Transactions on Multimedia Computing,
  Communications, and Applications (TOMM)}} \bibinfo{volume}{17},
  \bibinfo{number}{3} (\bibinfo{year}{2021}), \bibinfo{pages}{pp.1--18}.
\newblock


\bibitem[\protect\citeauthoryear{Zhang, Peng, and Yuan}{Zhang
  et~al\mbox{.}}{2018b}]%
        {ZhangPY18}
\bibfield{author}{\bibinfo{person}{Jian Zhang}, \bibinfo{person}{Yuxin Peng},
  {and} \bibinfo{person}{Mingkuan Yuan}.} \bibinfo{year}{2018}\natexlab{b}.
\newblock \showarticletitle{Unsupervised Generative Adversarial Cross-Modal
  Hashing}. In \bibinfo{booktitle}{\emph{Proceedings of the Thirty-Second
  {AAAI} Conference on Artificial Intelligence, (AAAI-18), the 30th innovative
  Applications of Artificial Intelligence (IAAI-18), and the 8th {AAAI}
  Symposium on Educational Advances in Artificial Intelligence (EAAI-18), New
  Orleans, Louisiana, USA, February 2-7, 2018}},
  \bibfield{editor}{\bibinfo{person}{Sheila~A. McIlraith} {and}
  \bibinfo{person}{Kilian~Q. Weinberger}} (Eds.). \bibinfo{publisher}{{AAAI}
  Press}, \bibinfo{address}{New Orleans, Louisiana, USA},
  \bibinfo{pages}{pp.539--546}.
\newblock
\urldef\tempurl%
\url{https://www.aaai.org/ocs/index.php/AAAI/AAAI18/paper/view/16746}
\showURL{%
\tempurl}


\bibitem[\protect\citeauthoryear{Zhang, Ma, Li, Huang, and Tian}{Zhang
  et~al\mbox{.}}{2016}]%
        {zhang2016pl}
\bibfield{author}{\bibinfo{person}{Liang Zhang}, \bibinfo{person}{Bingpeng Ma},
  \bibinfo{person}{Guorong Li}, \bibinfo{person}{Qingming Huang}, {and}
  \bibinfo{person}{Qi Tian}.} \bibinfo{year}{2016}\natexlab{}.
\newblock \showarticletitle{PL-Ranking: A Novel Ranking Method for Cross-Modal
  Retrieval}. In \bibinfo{booktitle}{\emph{Proceedings of the 24th ACM
  International Conference on Multimedia}} (Amsterdam, The Netherlands)
  \emph{(\bibinfo{series}{MM '16})}. \bibinfo{publisher}{Association for
  Computing Machinery}, \bibinfo{address}{New York, NY, USA},
  \bibinfo{pages}{pp.1355–1364}.
\newblock
\showISBNx{9781450336031}
\urldef\tempurl%
\url{https://doi.org/10.1145/2964284.2964336}
\showDOI{\tempurl}


\bibitem[\protect\citeauthoryear{Zhang, Dong, Du, Wu, Luo, and Yang}{Zhang
  et~al\mbox{.}}{2018a}]%
        {zhang2018collaborative}
\bibfield{author}{\bibinfo{person}{Xiang Zhang}, \bibinfo{person}{Guohua Dong},
  \bibinfo{person}{Yimo Du}, \bibinfo{person}{Chengkun Wu},
  \bibinfo{person}{Zhigang Luo}, {and} \bibinfo{person}{Canqun Yang}.}
  \bibinfo{year}{2018}\natexlab{a}.
\newblock \showarticletitle{Collaborative subspace graph hashing for
  cross-modal retrieval}. In \bibinfo{booktitle}{\emph{Proceedings of the 2018
  ACM on International Conference on Multimedia Retrieval}}.
  \bibinfo{publisher}{{ACM}}, \bibinfo{address}{Yokohama, Japan},
  \bibinfo{pages}{pp.213--221}.
\newblock
\urldef\tempurl%
\url{https://doi.org/10.1145/3206025.3206042}
\showDOI{\tempurl}


\bibitem[\protect\citeauthoryear{Zhen, Hu, Peng, Goh, and Zhou}{Zhen
  et~al\mbox{.}}{2020}]%
        {zhen2020deep}
\bibfield{author}{\bibinfo{person}{Liangli Zhen}, \bibinfo{person}{Peng Hu},
  \bibinfo{person}{Xi Peng}, \bibinfo{person}{Rick Siow~Mong Goh}, {and}
  \bibinfo{person}{Joey~Tianyi Zhou}.} \bibinfo{year}{2020}\natexlab{}.
\newblock \bibinfo{title}{Deep multimodal transfer learning for cross-modal
  retrieval}.
\newblock , \bibinfo{numpages}{13}~pages.
\newblock


\bibitem[\protect\citeauthoryear{Zhen, Hu, Wang, and Peng}{Zhen
  et~al\mbox{.}}{2019}]%
        {zhen2019deep}
\bibfield{author}{\bibinfo{person}{Liangli Zhen}, \bibinfo{person}{Peng Hu},
  \bibinfo{person}{Xu Wang}, {and} \bibinfo{person}{Dezhong Peng}.}
  \bibinfo{year}{2019}\natexlab{}.
\newblock \showarticletitle{Deep Supervised Cross-Modal Retrieval}. In
  \bibinfo{booktitle}{\emph{{IEEE} Conference on Computer Vision and Pattern
  Recognition, {CVPR} 2019, Long Beach, CA, USA, June 16-20, 2019}}.
  \bibinfo{publisher}{Computer Vision Foundation / {IEEE}},
  \bibinfo{address}{Long Beach, CA, USA}, \bibinfo{pages}{pp.10394--10403}.
\newblock
\urldef\tempurl%
\url{https://doi.org/10.1109/CVPR.2019.01064}
\showDOI{\tempurl}


\bibitem[\protect\citeauthoryear{Zhou, Wang, Fang, Bui, and Berg}{Zhou
  et~al\mbox{.}}{2018}]%
        {zhou2018visual}
\bibfield{author}{\bibinfo{person}{Yipin Zhou}, \bibinfo{person}{Zhaowen Wang},
  \bibinfo{person}{Chen Fang}, \bibinfo{person}{Trung Bui}, {and}
  \bibinfo{person}{Tamara~L. Berg}.} \bibinfo{year}{2018}\natexlab{}.
\newblock \showarticletitle{Visual to Sound: Generating Natural Sound for
  Videos in the Wild}. In \bibinfo{booktitle}{\emph{2018 {IEEE} Conference on
  Computer Vision and Pattern Recognition, {CVPR} 2018, Salt Lake City, UT,
  USA, June 18-22, 2018}}. \bibinfo{publisher}{Computer Vision Foundation /
  {IEEE} Computer Society}, \bibinfo{address}{Salt Lake City, UT, USA},
  \bibinfo{pages}{pp.3550--3558}.
\newblock
\urldef\tempurl%
\url{https://doi.org/10.1109/CVPR.2018.00374}
\showDOI{\tempurl}


\bibitem[\protect\citeauthoryear{Zhu, Ngo, Chen, and Hao}{Zhu
  et~al\mbox{.}}{2019}]%
        {Zhu_2019_CVPR}
\bibfield{author}{\bibinfo{person}{Bin Zhu}, \bibinfo{person}{Chong{-}Wah Ngo},
  \bibinfo{person}{Jingjing Chen}, {and} \bibinfo{person}{Yanbin Hao}.}
  \bibinfo{year}{2019}\natexlab{}.
\newblock \showarticletitle{{R2GAN:} Cross-Modal Recipe Retrieval With
  Generative Adversarial Network}. In \bibinfo{booktitle}{\emph{{IEEE}
  Conference on Computer Vision and Pattern Recognition, {CVPR} 2019, Long
  Beach, CA, USA, June 16-20, 2019}}. \bibinfo{publisher}{Computer Vision
  Foundation / {IEEE}}, \bibinfo{address}{Long Beach, CA, USA},
  \bibinfo{pages}{pp.11477--11486}.
\newblock
\urldef\tempurl%
\url{https://doi.org/10.1109/CVPR.2019.01174}
\showDOI{\tempurl}


\end{thebibliography}
\end{document}